\documentclass[aps,prb,preprint,onecolumn,amsmath,amssymb,superscriptaddress,eqsecnum,floatfix,scrartcl]{revtex4-1}
\pdfoutput=1      
\usepackage{amssymb}  
\usepackage{color,soul,colortbl} 
\usepackage{graphicx}  
\usepackage[normalem]{ulem}
\usepackage{subcaption}
\usepackage{mathrsfs}
\usepackage[pdftex,pdfusetitle,bookmarks=true,colorlinks=true,citecolor=blue,urlcolor=blue,linkcolor=magenta]{hyperref}
\usepackage{hypcap}
\usepackage{multirow}
\usepackage{float}
\usepackage{bm}
\usepackage{braket}
\usepackage{array}

\newcolumntype{x}[1]{>{\centering\arraybackslash\hspace{0pt}}p{#1}}

\definecolor{OliveGreen}{cmyk}{0.64, 0, 0.95, 0.40}
\definecolor{Gray}{gray}{0.9}

\begin{document}

\title{
Entanglement Spectrum as a diagnostic of chirality of Topological Spin Liquids: Analysis of an $\mathrm{SU}(3)$ PEPS}

\author{Mark J.~Arildsen}
\email{marildse@sissa.it}
\affiliation{Department of Physics, University of California, Santa Barbara, California 93106, USA}
\affiliation{SISSA --- International School for Advanced Studies and INFN, via Bonomea 265, 34136 Trieste, Italy}

\author{Ji-Yao Chen}
\affiliation{Guangdong Provincial Key Laboratory of Magnetoelectric Physics and Devices, Center for Neutron Science and Technology, School of Physics, Sun Yat-sen University, Guangzhou 510275, China}

\author{Norbert Schuch}
\affiliation{University of Vienna, Faculty of Mathematics,
Oskar-Morgenstern-Platz 1, 1090 Wien, Austria}
\affiliation{University of Vienna, Faculty of Physics,
Boltzmanngasse 5, 1090 Wien, Austria}

\author{Andreas W.~W.~Ludwig}
\affiliation{Department of Physics, University of California, Santa Barbara, California 93106, USA}

\date{\today} 
  
\begin{abstract}
    We address the key question of representation of chiral topological quantum states in (2+1) dimensions (i.e., with non-zero chiral central charge) by Projected Entangled Pair States (PEPS). A noted result (due to Wahl, Tu, Schuch, and Cirac [Phys.\ Rev.\ Lett.\ \textbf{111}, 236805 (2013)], and Dubail and Read [Phys.\ Rev.\ B \textbf{92}, 205307 (2015)]) says that this is possible for non-interacting fermions, but the answer is as yet unknown for interacting systems. Characteristic counting of degeneracies of low-lying states in the entanglement spectrum (ES) at fixed transverse momentum of bipartitioned long cylinders (``Li-Haldane counting’’) provides often-used supporting evidence for chirality. However, non-chiral PEPS (with zero chiral central charge), yet with strong breaking of time-reversal and reflection symmetries, with invariance under the product of these two operations (i.e., ``apparently'' chiral states), are known whose low-lying ES exhibits the same Li-Haldane counting as a chiral state in certain topological sectors [Kure\v{c}i\'c, Vanderstraeten, and Schuch, Phys.\ Rev.\ B \textbf{99}, 045116 (2019);  Arildsen, Schuch, and Ludwig, Phys.\ Rev.\ B \textbf{108}, 245150 (2023)]. In the present work, we identify a distinct indicator and hallmark of chirality in the ES of PEPS with global $\mathrm{SU}(3)$ symmetry: the splittings of conjugate irreps. We prove that in the ES of the chiral states conjugate irreps are exactly degenerate, because the operators that would split them [related to the cubic Casimir invariant of $\mathrm{SU}(3)$] are forbidden. By contrast, in the ES of non-chiral states, conjugate splittings are demonstrably non-vanishing. Such a diagnostic provides an unambiguous  and powerful tool  to distinguish chiral and non-chiral topological states in (2+1) dimensions via their entanglement spectra.
\end{abstract}

\maketitle 

\tableofcontents

\vskip 1cm

\section{Introduction}
\label{sec:intro}

The construction of the wavefunctions of chiral topological phases in (2+1) dimensions is an area of active research interest. These quantum states, which lack time-reversal symmetry, include the fractional quantum Hall states and chiral spin liquids, and have gapless topologically-protected boundaries with universal behavior determined by (1+1)D chiral conformal field theories (CFTs) characteristic to the topological order of the bulk.\cite{Halperin1982,Witten1989,Wen1990,Nayak2008} 
The entanglement entropy of such topological states satisfies an area law.\cite{Srednicki1993} States that satisfy such an entanglement entropy area law by construction can be built with the tensor network method of Projected Entangled Pair States (PEPS), which can additionally encode the symmetries of the state at the level of the local PEPS tensor.\cite{Verstraete2006,schuch:peps-sym}

In the present paper we address the important and fundamental question of whether PEPS wavefunctions can describe (2+1)-dimensional {\it chiral} topological quantum states. A famous result\cite{Dubail2015,Wahl2013} states that in the case of {\it non-interacting fermions,} PEPS wavefunctions {\it can} describe corresponding chiral non-interacting (2+1)-dimensional topological phases.
\footnote{A Hamiltonian possessing such a wave function of non-interacting fermions as a ground state turns out to be required to be either long-ranged, or gapless if local, a statement referred to as the ``no-go theorem''.}
But the question of whether PEPS wave functions are capable of describing {\it interacting} (2+1)-dimensional topological phases is open at present. The relevance for this question of the entanglement spectrum, the subject of this paper, is that the chiral nature of a (2+1)-dimensional quantum state is directly reflected in the entanglement spectrum (ES), in the form of data arising from a {\it chiral} (1+1)-dimensional conformal field theory (CFT), which is the same CFT as that characterizing a physical boundary of the chiral (2+1)-dimensional quantum state.
(We stress that, nevertheless, the {\it entanglement} spectrum is in general not the same as the spectrum of the CFT appearing at a physical boundary of the chiral (2+1)-dimensional quantum state, while both share the same ``Li-Haldane'' degeneracy counting of states occurring at fixed momentum.\cite{ArildsenLudwig2022})

Many numerical studies\cite{poilblanc:kl-peps-1,Poilblanc2016,Poilblanc2017,Hackenbroich2018,Chen2018,Chen2020} have had success constructing {\it interacting} PEPS wave functions that possess distinctive characteristics of chiral topological states, including such low-lying entanglement spectra with levels that exhibit the characteristic degeneracies of Li-Haldane state-counting\cite{Li2008} at fixed momentum of a chiral state.
However, it turns out that it is also possible for the low-lying entanglement spectrum of a {\it non-chiral} state to exhibit, in some topological sectors, precisely the same Li-Haldane counting of degeneracies as that of a {\it chiral} state, as in the case of the interacting PEPS discussed in Refs.~\onlinecite{Kurecic2019,ArildsenSchuchLudwig2022}.
It is thus of great interest to have a direct and unambiguous indicator that would identify interacting {\it chiral} topological PEPS, and would also in practice be accessible from a PEPS wave function.

In the present work, we consider a PEPS~\cite{Chen2020} with global $\mathrm{SU}(3)$ symmetry whose low-lying entanglement spectrum exhibits Li-Haldane counting of degeneracies at fixed momentum characteristic of the chiral topological $\mathrm{SU}(3)$ spin liquid described by chiral level-one $\mathrm{SU}(3)$ Chern-Simons theory. 
(We note that such chiral (2+1)-dimensional $\mathrm{SU}(3)$ and more general $\mathrm{SU}(N)$ spin liquids have been discussed in Refs.~\onlinecite{Gorshkov2010,Taie2012,Cappellini2014,Cazalilla2014,Pagano2014,Scazza2014,Zhang2014,Hofrichter2016,Nataf2016,Ozawa2018,Taie2022} and others in the setting of cold atom systems.)
In the context of this chiral level-one $\mathrm{SU}(3)$ spin liquid PEPS, we demonstrate in the present paper an indicator of the kind mentioned above, which goes beyond the Li-Haldane counting alone, to reveal chirality directly in the accessible, low-energy, finite-size entanglement spectrum: the degeneracy of conjugate pairs of $\mathrm{SU}(3)$ irreps in the entanglement spectrum.

This result comes from understanding the finite-size splittings of those degeneracies of the levels of the entanglement spectrum (i.e., of the spectrum of the entanglement Hamiltonian, at fixed momentum) that would be present in the spectrum of the Hamiltonian of the chiral conformal field theory taken on its own. 
In the case of chiral $\mathrm{SU}(2)$ level-one and level-two spin liquid PEPS, two of the authors have shown in previous work (Ref.~\onlinecite{ArildsenLudwig2022}) that low-lying entanglement spectra exhibit finite-size splitting structure that is precisely compatible with the underlying chiral topological quantum state, as manifested in the presence of corresponding terms (representing ``chiral conservation laws'') in the entanglement Hamiltonian, which can be identified as the source of the splittings.
In this work, we demonstrate that this result can be extended to chiral $\mathrm{SU}(3)$ PEPS by an analysis of the splittings of the entanglement spectrum of the chiral $\mathrm{SU}(3)$ spin liquid PEPS of Chen {\it et al.}~in Ref.~\onlinecite{Chen2020}.

Notably, however, we {\it moreover} observe exact degeneracies of conjugate states in the low-lying entanglement spectra reported in Ref.~\onlinecite{Chen2020}, which are precisely consistent with the exclusion of certain of the chiral conservation laws due to symmetry.
These exact degeneracies are in fact a property of the entanglement spectrum at arbitrarily high entanglement energy: A key result of the present paper is a proof that in the {\it chiral} entanglement Hamiltonian the only conservation laws that can be responsible for the splittings of conjugate representations are exactly forbidden by the defining symmetries of the topological states under consideration which are global $\mathrm{SU}(3)$ symmetry and the combined action $\mathcal{RT}$ of spatial reflection ${\cal R}$ and time-reversal ${\cal T}$.
By contrast, in another PEPS, invariant under the same $\mathrm{SU}(3)$ symmetry group and with the same symmetry properties under ${\cal T}$, ${\cal R}$ and ${\cal R T}$, and which was analyzed in Refs.~\onlinecite{Kurecic2019} and \onlinecite{ArildsenSchuchLudwig2022} and found to be {\it non-chiral}, the same degeneracy of conjugate states is not present, while (as already mentioned above) some sectors nevertheless precisely exhibit Li-Haldane counting of a chiral $\mathrm{SU}(3)$-level-one spin-liquid in the numerically accessible low-lying entanglement spectrum.
The lack of degeneracy of conjugate states in the low-energy finite-size entanglement spectra of Abelian $\mathrm{SU}(3)$ spin liquid states thus constitutes a direct indicator and a hallmark of a non-chiral state.

To obtain this result, we first begin by briefly reviewing some key features of the chiral $\mathrm{SU}(3)$ spin liquid PEPS from Ref.~\onlinecite{Chen2020} that we consider. This is done in Sec.~\ref{sec:peps}. Then, in Sec.~\ref{sec:wzw}, we briefly review the structure of the (1+1)D chiral $\mathrm{SU}(3)$-level-one [$\mathrm{SU}(3)_1$] Wess-Zumino-Witten CFT, which is the same as the chiral CFT that appears at a physical boundary of a chiral (2+1)D $\mathrm{SU}(3)$-level-one Chern-Simons topological field theory.\cite{Witten1989,Knizhnik1984}
We then review in Sec.~\ref{sec:boundarystate} the conformal boundary state description of the entanglement spectrum that guides our understanding of numerically observed low-lying entanglement spectra, a further development of the logic presented in Refs.~\onlinecite{ArildsenLudwig2022} and \onlinecite{Qi2012}. 
From this point of view, the splittings of the spectrum can be understood as the consequence of a Generalized Gibbs Ensemble of the conserved quantities (``conservation laws'') of the chiral CFT, in this case the chiral $\mathrm{SU}(3)_1$ CFT, that are allowed by the symmetries of the system. In Sec.~\ref{sec:conserved}, we discuss which conserved quantities these are: that is, we describe the corresponding set of integrals of boundary operators in the CFT representing these conservation laws, and then establish which are actually allowed to contribute to the GGE consistent with the global $\mathrm{SU}(3)$ symmetry and the discrete symmetries present. We then use a linear combination of the lowest-dimensional allowed conserved quantities appearing in the entanglement Hamiltonian to successfully perform fits to the low-lying entanglement spectra of Ref.~\onlinecite{Chen2020} in all available sectors: these results are presented in Sec.~\ref{sec:results}. 
Finally, in Sec.~\ref{sec:comparison}, we make the comparison with the non-chiral $\mathrm{SU}(3)$ PEPS of Refs.~\onlinecite{Kurecic2019,ArildsenSchuchLudwig2022}, which reveals the splitting of conjugate representations as a powerful tool to distinguish entanglement spectra of chiral and non-chiral topological states in (2+1) dimensions.

\section{Chiral $\mathrm{SU}(3)$ Spin Liquid PEPS}
\label{sec:peps}

\begin{figure}[H]
	\centering
	\includegraphics[scale=1.6]{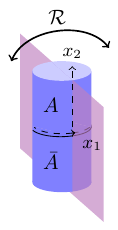}
	\caption{
            The spatial reflection $\mathcal{R}$ that together with time reversal $\mathcal{T}$ forms the $\mathcal{RT}$ transformation is depicted on a truncated representation of an infinite cylinder here. $\mathcal{R}$ is the reflection of the cylinder across the violet plane, which maps the upper half-cylinder $A$ and the lower half-cylinder $\bar{A}$ to themselves. This leaves the entanglement cut of the cylinder (shown as the solid circle at $x_2 = 0$) invariant. 
    }
	\label{fig:rtcylinder}
\end{figure}

The chiral $\mathrm{SU}(3)$ spin liquid PEPS we analyze in this work was first described by two of the authors of the present work (and collaborators) in Ref.~\onlinecite{Chen2020}. Ref.~\onlinecite{Chen2020} generalizes the results of Mambrini {\it et al.}~in Ref.~\onlinecite{Mambrini2016}. In that work, Mambrini {\it et al.}~classify the $\mathrm{SU}(2)$-symmetric PEPS on the square lattice, i.e., with $C_{4v}$ symmetry. The Supplementary Material of Ref.~\onlinecite{Chen2020} contains a similar classification for $\mathrm{SU}(3)$ PEPS with $C_{4v}$ symmetry. The chiral $\mathrm{SU}(2)$-level-one spin liquid of Poilblanc {\it et al.}~in Ref.~\onlinecite{poilblanc:kl-peps-1} is constructed from the basis given by tensors that transform under $C_{4v}$ and time-reversal in a way that leaves them invariant under $\mathcal{RT}$ symmetry, the combination of time reversal ($\mathcal{T}$) followed by a spatial reflection ($\mathcal{R}$). The spatial part ($\mathcal{R}$) of this transformation is depicted on a cylinder in Fig.~\ref{fig:rtcylinder}. In Ref.~\onlinecite{Chen2020}, the authors write down a similarly symmetric basis of PEPS tensors that serves as a variational ansatz. The properties of these tensors guarantee the global $\mathrm{SU}(3)$ symmetry of the PEPS, as well as the appropriate discrete $\mathcal{RT}$ symmetry. This ansatz can then be optimized with a CTMRG
\footnote{CTMRG = Corner Transfer Matrix Renormalization Group} 
method, minimizing the energy of the state (i.e., the expectation value, on a region embedded in the infinite plane, of a particular Hamiltonian with three-site short-ranged $\mathrm{SU}(3)$-symmetric interactions that is shown by exact diagonalization to host a chiral $\mathrm{SU}(3)$ spin liquid phase).\cite{Chen2020}

The authors of Ref.~\onlinecite{Chen2020} compute the entanglement spectrum of this PEPS on a half-infinite cylinder of circumference $N_v = 6$. We will consider this data below.
\footnote{This spectrum is shown in Fig.~3 of Ref.~\onlinecite{Chen2020}.}
Crucially, as the authors note, the entanglement spectra exhibit countings of $\mathrm{SU}(3)$ irreps, i.e., degeneracies in momentum of those irreps, that at the lowest levels of the entanglement spectrum are characteristic of the chiral $\mathrm{SU}(3)_1$ CFT. In particular, in Ref.~\onlinecite{Chen2020}, the entanglement spectrum of the PEPS block-diagonalizes into sectors of three different total $\mathbb{Z}_3$ charges $Q$ for the state of virtual links on the boundary of the half-infinite cylinder. These sectors are denoted by $Q = 0, \pm 1$. The lowest levels of the entanglement spectrum in the $Q = 0$ sector exhibit degeneracies characteristic of the $\bm{1}$ (singlet/trivial representation of $\mathrm{SU}(3)$) primary sector of the chiral $\mathrm{SU}(3)_1$ CFT, while the $Q = \pm 1$ sectors' entanglement spectra exhibit degeneracies characteristic of the $\bm{3}$ and $\overline{\bm{3}}$ (fundamental and anti-fundamental representations of $\mathrm{SU}(3)$) primary sectors of the chiral $\mathrm{SU}(3)_1$ CFT. (These sectors will be explained in more detail in the following section reviewing the chiral $\mathrm{SU}(3)_1$ CFT.) This is the starting point for the analysis to follow.

\section{Chiral $\mathrm{SU}(3)_1$ Wess-Zumino-Witten Theory}
\label{sec:wzw}

We first briefly review the basics of the chiral $\mathrm{SU}(3)_1$ Wess-Zumino-Witten (WZW) conformal field theory.\cite{Knizhnik1984,Witten1984,DiFrancesco1997}  
This theory has three primary sectors, associated to primary states arranged into each of the three lowest-dimensional representations of $\mathrm{SU}(3)$: $\bm{1}$, $\bm{3}$, and $\overline{\bm{3}}$.
The main ingredient in building up the Hilbert space of the chiral $\mathrm{SU}(3)_1$ WZW theory from the primary states is the Noether current of the global $\mathrm{SU}(3)$ symmetry $J^a(x)$ where $a = 1,\ldots,8$, and where we place the theory on a circle of circumference $\ell$, so $x$ is the periodic spatial coordinate.
The energy-momentum tensor $T(x)$ of the CFT can be built from the $J^a(x)$ as expressed in the Sugawara form
\begin{equation}
\label{eq:sugawara}
T(x) = \frac{1}{k+3}\sum_{a=1}^8 (J^a J^a)(x).
\end{equation}
The periodicity of the spatial coordinate $x$ motivates mode expansions of $J^a(x)$ and $T(x)$:
\begin{equation}
\label{eq:modeexpansion}
J^a(x) = \frac{2\pi}{\ell}\sum_{n=-\infty}^\infty J^a_{n}e^{2\pi inx/\ell}\text{ and }T(x) =\left(\frac{2\pi}{\ell}\right)^2 \left(-\frac{c}{24}+\sum_{n=-\infty}^\infty L_{n}e^{2\pi inx/\ell}\right),
\end{equation}
where the central charge $c = 2$ for the $\mathrm{SU}(3)_1$ case. 
The $J^a_n$ can then be used to build up the Hilbert space of the chiral $\mathrm{SU}(3)_1$ WZW theory from the primary states by writing states of the form (not necessarily distinct)
\begin{equation}
\label{eq:jbasis}
J^{a_1}_{-n_1}\cdots J^{a_m}_{-n_m}\ket{r,\sigma},
\end{equation}
where $\ket{r,\sigma}$ is a primary state with representation $r \in \{\bm{1},\bm{3},\overline{\bm{3}}\}$ and $\sigma$ the specific state in the representation, and the $n_i$ are positive integers. 
We define the level of the state to be $n = \sum_i n_i$. While the form shown does not make it manifest, the states of Eq.~\eqref{eq:jbasis} at a given level can be organized into representations of $\mathrm{SU}(3)$. The number of each kind of $\mathrm{SU}(3)$ representation present in the Hilbert space at each level constitutes a characteristic signature of the chiral $\mathrm{SU}(3)_1$ WZW theory. The states of Eq.~\eqref{eq:jbasis} are all eigenvalues of $L_0$, and we have 
\begin{equation}
    L_0 \left(J^{a_1}_{-n_1}\cdots J^{a_m}_{-n_m}\ket{r,\sigma} \right) = \left(h_r + n\right)\left(J^{a_1}_{-n_1}\cdots J^{a_m}_{-n_m}\ket{r,\sigma} \right),
\end{equation}
where $h_r$ is the conformal weight of the primary state, with $h_{\bm{1}} = 0$ and $h_{\bm{3}} = h_{\bar{\bm{3}}} = 1/3$. 
We can derive a Hamiltonian for this chiral theory on the edge of the cylinder from the integral of the energy-momentum tensor $T(x)$, yielding
\begin{equation}
    \label{eq:hcft}
    H_{\text{CFT}} = v P = \frac{v}{2\pi}\int_0^{\ell} T(x)dx = \frac{2\pi v}{\ell} \left( L_0 - \frac{c}{24}\right),
\end{equation}
where $v$ is the velocity of the edge states, and $P$ is the momentum.
All of the states of Eq.~\eqref{eq:jbasis} are eigenstates of $H_{\text{CFT}}$ by virtue of being eigenstates of $L_0$. As states at the same level have the same $P$ and $H_{\text{CFT}}$ eigenvalue, states in all of the $\mathrm{SU}(3)$ representations at a given level will be degenerate both in momentum and in energy.

\section{Conformal Boundary State Approach to the Entanglement Spectrum}
\label{sec:boundarystate}

We consider entanglement spectra of a topological phase described by chiral (2+1)D $\mathrm{SU}(3)$-level-one Chern-Simons theory, on an infinite cylinder bipartitioned by a circumferential cut of finite size $\ell$. In the lower energy levels of these entanglement spectra, we expect to see (and the numerical study considered here confirms) a left-moving chiral branch. This is related, as we will explain, to the left-moving chiral CFT that would be present on the edge were the infinite cylinder to be physically bipartitioned along the cut. Such a theory would be the chiral $\mathrm{SU}(3)_1$ WZW CFT (defined on a circle) of Sec.~\ref{sec:wzw}. Eigenstates in the lower levels of the entanglement Hamiltonian form $\mathrm{SU}(3)$ representations, and in the lower-entanglement energy part of the spectrum, the number of each kind of representation degenerate at each value of the momentum and proximal in entanglement energy is consistent with the numbers of each kind of representation at the corresponding level of the chiral $\mathrm{SU}(3)_1$ WZW theory. 

The degeneracy in energy of the states at each level, however, is plainly broken in these entanglement spectra. To explain this, we follow the approach in Ref.~\onlinecite{ArildsenLudwig2022}, by two of the authors, who understand the splitting of these degeneracies in the entanglement spectrum from a conformal boundary state approach. In the same manner as Ref.~\onlinecite{Qi2012}, they relate the ground state of the (2+1)D chiral topological state on a cylindrical geometry to a conformally invariant fixed point boundary state on the entanglement cut, describing a (conformally invariant) boundary condition on the bulk version of the underlying CFT; however, they incorporate the insights obtained in Ref.~\onlinecite{Dubail2012}, and Ref.~\onlinecite{Cardy2016} in the context of quantum quenches, to see that this relation must take into account the full Generalized Gibbs Ensemble (GGE) of the conserved integrals of irrelevant local boundary operators. These boundary operators must be the (coinciding) boundary limits of both left- and right-moving (i.e., holomorphic and anti-holomorphic) chiral bulk operators $\Phi_i$ and $\overline{\Phi}_i$ of the CFT, respectively, which satisfy $\Phi_i(x) = \overline{\Phi}_i(x)$ on the boundary. 
When we then compute the left-moving reduced density matrix $\rho_{L,a}$ obtained by tracing out right-moving degrees of freedom of the (bulk) CFT, restricting to the topological sector of label $a$ (where $P_a$ denotes projection into said sector), we obtain
\begin{equation}
\label{eq:realrhol}
    \rho_{L,a} \propto P_a e^{-\beta H_L} \prod_i e^{-\beta_i  \int \Phi_i(x) dx} P_a,
\end{equation}
where $H_L$ is the Hamiltonian of the left-moving chiral (1+1)-dimensional CFT.
$\beta$ and the $\beta_i$ are effective inverse temperature parameters associated to the corresponding conserved integrals of the GGE, and the $\Phi_i(x)$ are irrelevant operators acting only on the left-moving CFT. 
We can then define the locally conserved quantities $H^{(i)}$ by $H^{(i)} = \frac{1}{2\pi} \int_0^\ell \Phi_i(x) dx$. 
Thus the overall entanglement Hamiltonian can be written in terms of $H_L$ and the other (left-moving) conserved quantities $H^{(i)}$ of the GGE as
\begin{equation}
    \label{eq:lincomb}
    H_{\text{entanglement}} -{\rm const.}= \beta H_{L} + \sum_{i=2}^\infty \beta_i H^{(i)},
\end{equation}
where we take $H^{(1)} = H_L$ and $\beta_1 = \beta$, and ``const.'' ensures the proper normalization of the entanglement Hamiltonian.
\footnote{See, e.g., the discussion regarding this constant term in Ref.~\onlinecite{ArildsenLudwig2022}.}
Eq.~\eqref{eq:lincomb} encapsulates the central concept of this work. After determining what the required conserved quantities $H^{(i)}$ (up to conformal dimension $\Delta \leq 6$) should be in Sec.~\ref{sec:conserved}, we use Eq.~\eqref{eq:lincomb}, with the $\beta_i$ as free parameters, to fit the data of the entanglement spectrum of the PEPS from Sec.~\ref{sec:peps}. 
Physically, the particular linear combination of GGE conservation laws appearing in the entanglement Hamiltonian, specified by the set of these parameters $\beta_i$, reflects the particular wave function within the topological phase on the surface of the cylinder in the entanglement spectrum. This analysis is carried out in Sec.~\ref{sec:results}.

\section{Determining the Conserved Quantities Present}
\label{sec:conserved}

The conserved quantities we expect to be present in the GGE described in the previous section must satisfy several criteria. They must commute with the chiral $\mathrm{SU}(3)_1$ WZW CFT Hamiltonian of Eq.~\eqref{eq:hcft}. The conserved quantities must also be invariant under global $\mathrm{SU}(3)$ symmetry. Further, the conserved quantities need to be invariant under certain discrete symmetries that preserve the entanglement cut. One such symmetry is the $\mathcal{RT}$ symmetry mentioned in Sec.~\ref{sec:peps}, and here implemented as the combination of time reversal ($\mathcal{T}$) followed by spatial reflection of the chiral topological state about a plane ($\mathcal{R}$) in such a way that the entanglement cut is mapped into itself.

The available conserved quantities from the chiral $\mathrm{SU}(3)_1$ WZW CFT that are invariant under the global $\mathrm{SU}(3)$ symmetry correspond to the $\mathrm{SU}(3)$ singlet descendants of the identity primary state sector $\ket{\bm{1}}$. These conserved quantities consist of integrals of normal-ordered combinations and derivatives of the energy-momentum tensor $T(x)$, which, like the quadratic Casimir invariant of $\mathrm{SU}(3)$, is a bilinear when written in the Sugawara form of Eq.~\eqref{eq:sugawara}, and the ``$W_3$-current'' current $W(x)$, which obeys\cite{Bouwknegt1993} 
\begin{equation}
    \label{eq:wcurrent}
    W(x) \propto \sum_{a,b,c =1}^8 d_{abc} (J^a(J^b J^c))(x),
\end{equation}
where $d_{abc}$ is a completely symmetric traceless tensor, and thus has the trilinear form of the {\it cubic} Casimir invariant of $\mathrm{SU}(3)$.\cite{Biedenharn1963} 
Importantly for what follows, the eigenvalue of the cubic Casimir invariant of $\mathrm{SU}(3)$ has the opposite sign on conjugate representations, while the eigenvalue of the quadratic Casimir invariant has the same sign on conjugate representations. This property extends to the conserved quantities constructed from $T(x)$ and $W(x)$, as is discussed in greater detail in Appendix \ref{app:calculation}.

Not all of these conserved quantities will satisfy the additional criteria imposed by invariance under discrete symmetries, however. In particular, the conserved quantities formed from the integrals of irrelevant operators of odd conformal dimension will not contribute, as these quantities turn out to be precisely those which are not even under the discrete $\mathcal{RT}$ symmetry.
See Appendix~\ref{app:calculation} and Appendix \ref{app:discrete}, which contain a more in-depth discussion of which conserved quantities should be excluded from our consideration on grounds of the $\mathcal{RT}$ symmetry.
Crucially, we have proven (in Appendix \ref{app:calculation}) that all conservation laws, at arbitrarily high operator dimension, whose eigenvalues have opposite sign on conjugate representations are odd under the discrete $\mathcal{RT}$ symmetry and are thus excluded from appearing in the entanglement Hamiltonian. It is this result that implies our finding of conjugate degeneracy as a necessary condition of chirality in these entanglement spectra.
Additionally, but less importantly, we find empirically from the fits of Sec.~\ref{sec:results} that the operator which will be denoted by $(T\partial W)(x)$, and which is of even operator dimension and does {\it not} lead to a splitting of conjugate states, happens to be excluded as well, very likely due to a related but different discrete symmetry. 

We list in Table \ref{table:operators} the full set of irrelevant operators of dimension $\Delta \leq 6$ that are integrated to give conserved quantities present in Eq.~\eqref{eq:lincomb}. The conserved quantities are written in terms of modes in Table \ref{table:modereps} in Appendix~\ref{app:calculation}.

\begin{table}[hbt]
	\centering
	\begin{tabular}{c|c|>{\columncolor{Gray}}c}
		$\Delta$ & $\Phi_i(x)$ included & $\Phi_i(x)$ excluded \\
		\hline
		2 & $T(x)$ & --- \\
		\hline
		3 & --- & $W(x)$ \\
		\hline
		4 & $(TT)(x)$ & --- \\
		\hline
		5 & --- & $(TW)(x)$ \\
		\hline
		6 & $(T(TT))(x)$, $(\partial T \partial T)(x)$, $(WW)(x)$ & $(T\partial W)(x)$ \\
		\hline
	\end{tabular}
\caption{An enumeration of the $\mathrm{SU}(3)$-invariant irrelevant operators $\Phi_i(x)$ of conformal dimensions $\Delta \leq 6$ (with operators that are total derivatives and thus trivially excluded not shown). The operators in the ``included'' column are used to fit the splittings of the numerical spectra within each descendant level for the chiral $\mathrm{SU}(3)_1$ WZW theory we consider. Those in the ``excluded'' column violate discrete symmetries of the system and therefore are excluded from our considerations.}
\label{table:operators}
\end{table}

\section{Results for the chiral $\mathrm{SU}(3)$ PEPS}
\label{sec:results}

\begin{figure}[H]
	\centering
	\includegraphics[scale=0.4]{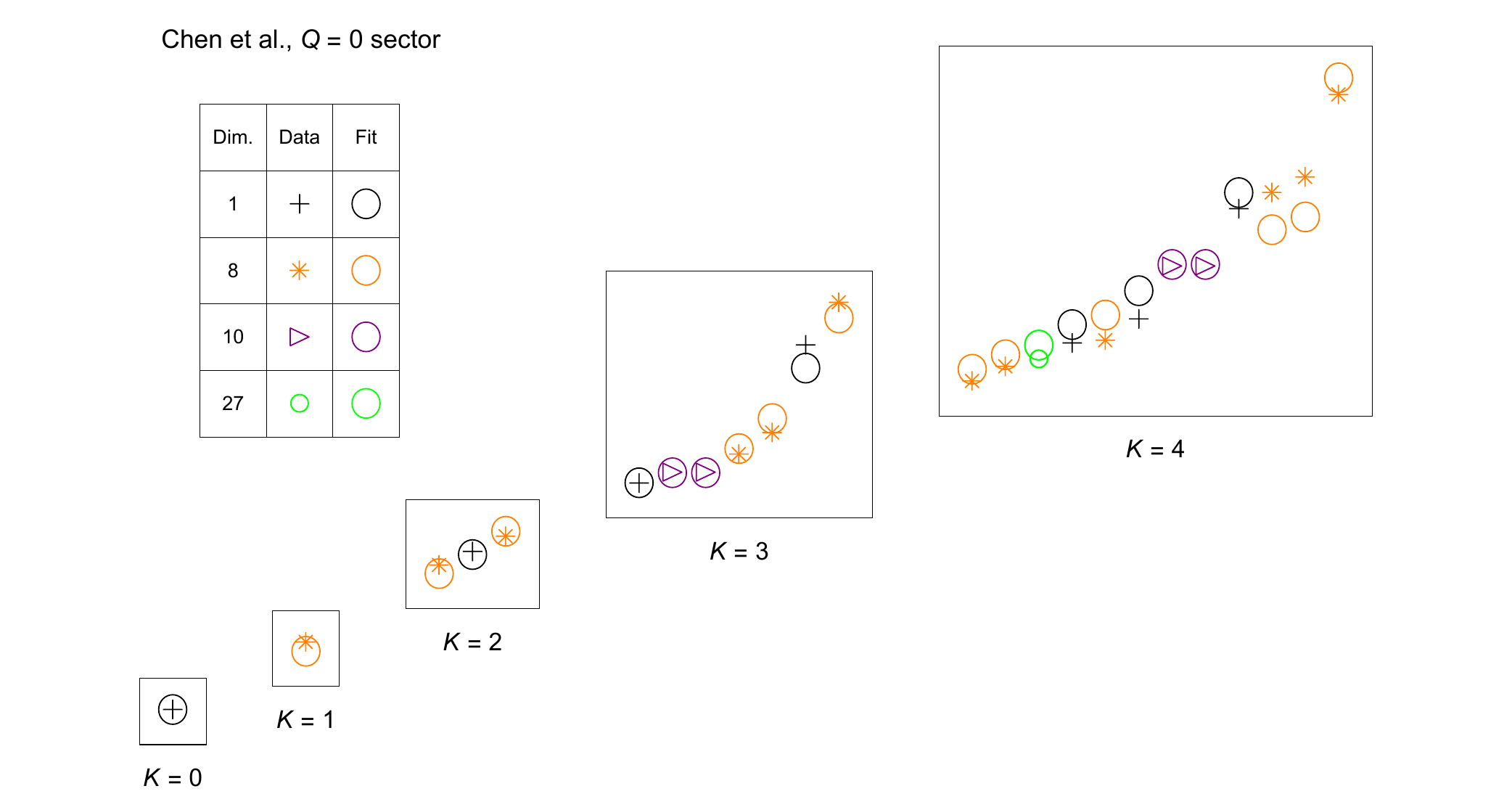}
	\caption{Fit of the lower levels of the $Q= 0$ entanglement spectrum, corresponding to the $\ket{\bm{1}}$ primary sector of the chiral $\mathrm{SU}(3)_1$ WZW theory. In this fit, 5 parameters, corresponding to the conserved integrals of the operators of the ``$\Phi_i(x)$ included'' column of Table \ref{table:operators}, are used to fit the 23 differences among the energies of the 24 $\mathrm{SU}(3)$ multiplets in the first 5 levels of the $\ket{\bm{1}}$ sectors.}
	\label{fig:chirales0fit}
\end{figure}

\begin{figure}[H]
	\centering
	\includegraphics[scale=0.4]{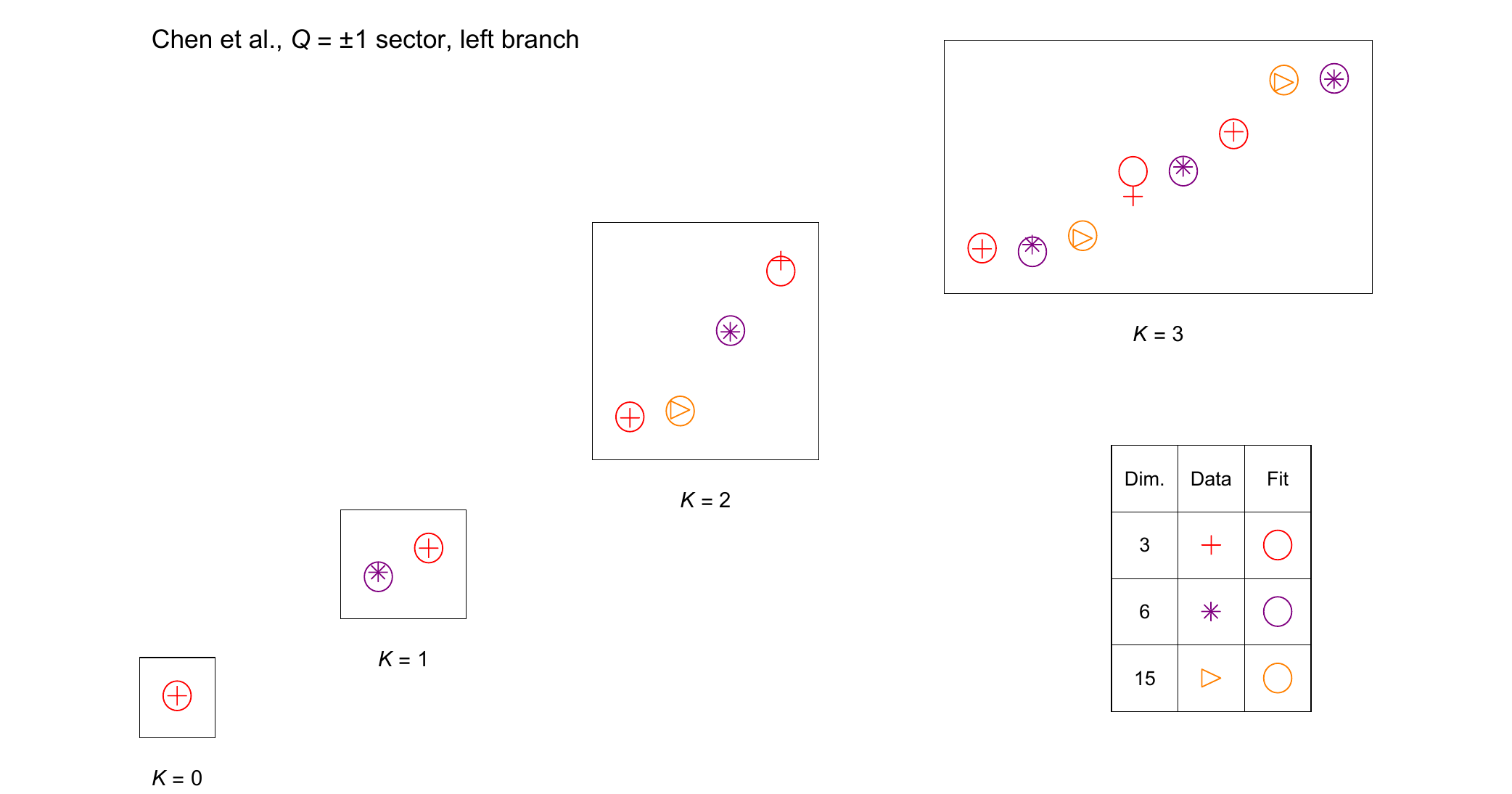}
	\caption{Fit of the left branch of the lower levels of the $Q=\pm 1$ entanglement spectrum, corresponding to the $\ket{\bm{3}}$ or $\ket{\overline{\bm{3}}}$ primary sectors of the chiral $\mathrm{SU}(3)_1$ WZW theory. In this fit, 5 parameters, corresponding to the conserved integrals of the operators of the ``$\Phi_i(x)$ included'' column of Table \ref{table:operators},  are used to fit the 14 differences among the energies of the 15 $\mathrm{SU}(3)$ multiplets in the first 4 levels of the $\ket{\bm{3}}$ and $\ket{\overline{\bm{3}}}$ sectors.}
	\label{fig:chirales1fit}
\end{figure}

We can then fit the numerical data of the lower levels of the entanglement spectrum produced from the PEPS of Sec.~\ref{sec:peps} to a truncation of Eq.~\eqref{eq:lincomb}. For these fits, we consider only the most relevant of the conservation laws $H^{(i)}$ of Eq.~\eqref{eq:lincomb}, which are the integrals of the currents listed in Table \ref{table:operators},
\footnote{Explicit descriptions of these integrals are given in terms of modes of the currents in Table \ref{table:modereps} in Appendix \ref{app:calculation}.}
and we find the corresponding parameters $\beta_i$ for Eq.~\eqref{eq:lincomb} that yield the closest entanglement spectrum to that of the PEPS for the lower levels being fit. We choose to fit the entanglement spectrum only up to the entanglement energy levels beyond which it becomes difficult to sort out from the numerical data which multiplets belong to which descendant level above the corresponding primary state in the chiral $\mathrm{SU}(3)_1$ WZW theory. 
These fits are shown in Figs.~\ref{fig:chirales0fit} and \ref{fig:chirales1fit}. 

Figure \ref{fig:chirales0fit} depicts the fit to the entanglement spectrum of the PEPS in the $Q = 0$ sector, which, as discussed above in Sec.~\ref{sec:peps}, contains Li-Haldane countings corresponding to the primary singlet/identity state sector $\ket{\bm{1}}$ of the chiral $\mathrm{SU}(3)_1$ CFT. We use 5 parameters corresponding to the $\beta_i$ of Eq.~\eqref{eq:lincomb} to fit the 23 differences among the 24 $\mathrm{SU}(3)$ multiplets of the first 5 levels of the $\ket{\bm{1}}$ primary state sector (Fig.~\ref{fig:chirales0fit}). 

Fitting the entanglement spectrum of the PEPS in the $Q = \pm 1$ sector is more complicated, as there are three chiral branches present in the data for each of the $Q=+1$ and the $Q=-1$ sector, where each branch appears to exhibit the characteristic Li-Haldane counting of the $\ket{\bm{3}}$ and $\ket{\overline{\bm{3}}}$ primary state sectors. We will refer to the branches with the primary state (lowest entanglement energy state) located at momenta $K \pmod{2\pi} = -\pi/3$, $\pi/3$, and $\pi$ as the left, center, and right branches, respectively. Fig.~\ref{fig:chirales1fit} depicts the fit to the entanglement spectrum in the left branch. We use 5 parameters corresponding to the same conserved quantities to fit the 14 differences among the energies of the 15 $\mathrm{SU}(3)$ multiplets in the first 4 levels of the $\ket{\bm{3}}$ and $\ket{\overline{\bm{3}}}$ sectors (Fig.~\ref{fig:chirales1fit}) found in the left branch of the entanglement spectrum of the PEPS. Further fits to the PEPS entanglement spectra can be found in Appendix \ref{sec:furtherfits}, while the parameters $\beta_i$ for those fits and the fits of Figs.~\ref{fig:chirales0fit} and \ref{fig:chirales1fit}, as well as a more detailed discussion of the fitting algorithm, can be found in Appendix \ref{app:params}. 

Looking at these fits, one immediate observation is that the conjugate $\bm{10}$ and $\overline{\bm{10}}$ representations in the sector of the $\mathrm{SU}(3)$ identity primary $\ket{\bm{1}}$, which appear as the pair of 10-dimensional representations in Fig.~\ref{fig:chirales0fit} (denoted by the purple triangle markers), have the same entanglement energy, to a very close approximation---in particular, much smaller than the scale of the other level splittings in the entanglement spectrum.
This is wholly consistent with our result that, by $\mathcal{RT}$ symmetry, the conserved quantities in the chiral $\mathrm{SU}(3)_1$ CFT exclude those which differentiate between conjugate multiplets---i.e., the conserved quantities that are integrals of operators of odd dimension (such as, e.g., $W(x)$ and $(TW)(x)$). (See Table \ref{table:operators}.) This consistency with the invariance of the allowed conserved quantities under conjugation of representations is also visible in the overall degeneracy of the lower levels of the $Q = 1$ and $Q = -1$ entanglement spectra, which contain data corresponding to the conjugate $\ket{\bm{3}}$ and $\ket{\overline{\bm{3}}}$ primary sectors.
The left-branch of the low-lying entanglement spectra of the levels of the $Q=+1$ and $Q=-1$ sectors is seen in Fig.~\ref{fig:chirales1fit}, as noted above. The $Q=+1$ and $Q=-1$ sectors are not displayed separately as their degeneracy renders them effectively identical. This will be discussed in more detail in Sec.~\ref{sec:comparison}.

It is not possible to understand the exclusion of the integral of $(T\partial W)(x)$, defined in Table \ref{table:operators}, in the same way, though. Like the other conserved quantities that are integrals of even-dimensional operators, it has the same value on multiplets of conjugate representations. Notably, however, our fitting approach can reveal that it, too, should be excluded, in the sense that many of the fits making use of all of the conserved quantities associated to the even-dimensional operators in Table \ref{table:operators} yield a value of the coefficient $\beta_7$ in Eq.~\eqref{eq:lincomb} associated to the integral of $(T\partial W)(x)$ that is several orders of magnitude smaller than the other $\beta_i$. (In fact, the normalized $\tilde{\beta}_7$ value, comparable to the normalized values $\tilde{\beta}_i$ of the other conservation laws displayed in Table \ref{table:su31fitparams}, is $\sim 10^{-8}$.)


\section{Comparison with Splittings in Non-Chiral $\mathrm{SU}(3)$ PEPS with Strong Time-Reversal and Parity Symmetry Breaking}
\label{sec:comparison}

The phenomenon described in the previous section, whereby conjugate states in the entanglement spectra depicted in Figs.~\ref{fig:chirales0fit}--\ref{fig:chirales1fit} are degenerate, is a necessary consequence of the exclusion, due to $\mathcal{RT}$ symmetry {\it in the chiral case}, of the odd dimensional operators [such as $W(x)$ and $(TW)(x)$] whose integrals would split the conjugate pairs. However, this is in contrast to the splitting between conjugate states that {\it does occur} in the case of the {\it non-chiral} PEPS, also in the presence of $\mathcal{RT}$ symmetry, analyzed in Refs.~\onlinecite{Kurecic2019} and \onlinecite{ArildsenSchuchLudwig2022}. In that PEPS, the entanglement spectra describe a non-chiral (or: ``doubled'', in the sense of possessing right ($R$) and left ($L$) moving sectors) theory as outlined in Ref.~\onlinecite{ArildsenSchuchLudwig2022}, a theory that has nine topological sectors, which descend from tensor products of the three primary states of each of the left- and right-moving chiral $\mathrm{SU}(3)_1$ WZW CFTs reviewed in Sec.~\ref{sec:wzw}. 

To understand the splitting between conjugate states in the non-chiral PEPS, we will first briefly review the result of Ref.~\onlinecite{ArildsenSchuchLudwig2022} on the structure of the entanglement Hamiltonian of that PEPS, and then we will discuss the essence of a perturbative approach to the splittings found in that entanglement Hamiltonian.

\subsection{Review of non-chiral PEPS}
The non-chiral PEPS of Ref.~\onlinecite{Kurecic2019} is an $\mathrm{SU}(3)$ topological spin liquid on the kagome lattice. It has the property that its entanglement spectrum is gapped, and possesses two chiral branches, a left-moving branch and a right-moving branch, of which the left-moving branch possesses a substantially greater velocity. 

\begin{figure}[H]
    \centering
    \includegraphics[scale=2]{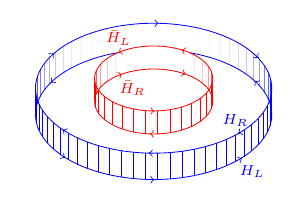}
	\caption{The boundaries of two cut cylinders are shown at the location of the cut. The outer, blue cylinder represents the chiral, $\mathrm{SU}(3)_1$ part of the double Chern-Simons theory, while the inner, red cylinder represents the anti-chiral, $\overline{\mathrm{SU}(3)_1}$ part. The direction of motion on each edge is depicted by arrows, and the corresponding boundary Hamiltonians are labeled. The vertical links shown between the boundaries of both the inner and outer cylinders represent the couplings between $H_L$ and $H_R$, and between $\bar{H}_L$ and $\bar{H}_R$. Note that this is distinct from the couplings on each side of the boundary between $H_L$ and $\bar{H}_R$, and between $H_R$ and $\bar{H}_L$, due to $\mathcal{H}_{\text{interaction}}$ and $\overline{\mathcal{H}}_{\text{interaction}}$, respectively.}
	\label{fig:boundarycylindergraphic}
\end{figure}

Ref.~\onlinecite{ArildsenSchuchLudwig2022} decomposes the structure of the low-lying entanglement Hamiltonian of the non-chiral PEPS as
\begin{equation}
\label{eq:htotal}
    H_{\text{total}} = H_L + \bar{H}_R + \mathcal{H}_{\text{interaction}} = H_{\text{doubled}} + \mathcal{H}_{\text{interaction}},
\end{equation}
where $H_L$ is the Hamiltonian of the left-moving chiral theory on the edge (that would appear on the cylinder physically cut at the entanglement cut), $\bar{H}_R$ is the right-moving chiral theory on the same edge (that would appear at the same physical cut), and $\mathcal{H}_{\text{interaction}}$ is the interaction between them. The decoupled sum of these two left- and right-moving chiral Hamiltonians is denoted by $H_{\text{doubled}} = H_L + \bar{H}_R$. For the opposite boundary of the cut cylinder, the entanglement Hamiltonian (in that case, obtained by tracing out the opposite half of the bipartitioned cylinder) is $\bar{H}_{\text{total}}$, which we can write as
\begin{equation}
    \bar{H}_{\text{total}} = \bar{H}_L + H_R + \bar{\mathcal{H}}_{\text{interaction}} = \bar{H}_{\text{doubled}} + \bar{\mathcal{H}}_{\text{interaction}}.
\end{equation}
This description can be represented graphically by the representation in Fig.~\ref{fig:boundarycylindergraphic}, which illustrates the two counter-propagating theories (in blue and red) found on each boundary of the cut cylinder.

In the entanglement spectrum of the non-chiral PEPS, we observe contributions in addition to the splittings arising from the conserved quantities arising from the purely chiral $\mathrm{SU}(3)_1$ theory discussed in previous sections. While those splittings ought indeed to still be manifest in the entanglement spectrum, the $\mathcal{H}_{\text{interaction}}$ term in Eq.~\eqref{eq:htotal} gives rise to additional contributions to the entanglement Hamiltonian that lead to a distinct pattern of splittings.

\subsection{Splittings from $\mathcal{H}_{\text{interaction}}$ in the non-chiral PEPS}
\label{sec:nonchiralsplit}

\begin{table}
\begin{tabular}{c|c|c}
--- & $\mathcal{H}^{(i)}$ respecting $\mathrm{SU}(3) \times \overline{\mathrm{SU}(3)}$ & $\mathcal{H}^{(i)}$ respecting $\mathrm{SU}(3)$ only \\
\hline
$\mathcal{H}^{(i)}$ that do \emph{not} split conjugate irreps & $T\overline{T}$, $\ldots$ & $J^a\overline{J}^a$, $\ldots$  \\
\hline
$\mathcal{H}^{(i)}$ that do split conjugate irreps & $T\overline{W}$, $W\overline{T}$, $\ldots$ & $J^a\overline{T}^a$, $T^a\overline{J}^a$, $(WJ^a)\overline{J}^a$, $J^a\overline{(WJ^a)}$, $\ldots$
\end{tabular}

\caption{Four categories of interaction terms that contribute to $\mathcal{H}_{\text{interaction}}$. This subsection, Sec.~\ref{sec:nonchiralsplit}, primarily explores the second column, consisting of terms that break the $\mathrm{SU}(3) \times \overline{\mathrm{SU}(3)}$ symmetry and thus split the irreps of diagonal $\mathrm{SU}(3)$ found in a given tensor product of states from the left- and right-moving chiral theories. Meanwhile, Sec.~\ref{sec:conjugate} addresses the bottom row, operators that split conjugate pairs of $\mathrm{SU}(3)$ irreps. The notation of adjacent left- and right-moving operators used in this table is illustrative of the interaction terms constructed as in the equations for specific $\mathcal{H}^{(i)}$ given in the main text, e.g. Eq.~\eqref{eq:jjbarperturbation} for $\mathcal{H}^{(1)}$.}
\label{table:interactionterms}
\end{table}

We can analyze the splittings in the non-chiral PEPS due to $\mathcal{H}_{\text{interaction}}$ by writing it as a linear combination of right ($R$)/left ($L$)-compound terms:
\begin{equation}
    \label{eq:hinteraction}
    \mathcal{H}_{\text{interaction}} = \sum_{i=1}^n \lambda_i\mathcal{H}^{(i)},
\end{equation}
where the $\lambda_i$ are real coefficients, and the $\mathcal{H}^{(i)}$ themselves will consist of zero-(total)-momentum combinations of modes of operators from the left- and right-moving chiral theories, as a perturbation to the decoupled Hamiltonian $H_L + \bar{H}_R$. We will describe the leading (in the renormalization group sense) such $\mathcal{H}^{(i)}$ below and in the following subsection. These terms will fill out the four sections of Table \ref{table:interactionterms}.

$\mathcal{H}^{(1)}$, the first example, is perhaps the clearest facet of the non-chiral (``doubled'') $\mathrm{SU}(3)$ entanglement spectrum that we are able to explain by a first-order perturbation. $\mathcal{H}^{(1)}$ gives rise to the splittings of representations of the diagonal (``global'') $\mathrm{SU}(3)$ symmetry that come from particular tensor products of representations in $\mathrm{SU}(3) \otimes \overline{\mathrm{SU}(3)}$. It is thus not possible to explain these splittings at all solely from our understanding of the
decoupled left- and right-moving chiral theories, as these splittings break the degeneracy imposed by $\mathrm{SU}(3) \otimes \overline{\mathrm{SU}(3)}$ down to its diagonal subgroup leaving only global $\mathrm{SU}(3)$ symmetry intact. Such splittings are seen to occur in the six sectors of $\mathrm{SU}(3)_1 \otimes \overline{\mathrm{SU}(3)}_1$ where the ``fast'' sector has a non-singlet, i.e., $\ket{\bm{3}}_L$ or $\ket{\overline{\bm{3}}}_L$ primary state. [The $L$ subscript denotes that the primary is in the ``fast'', left-moving ($L$) chiral branch. An $R$ subscript denotes a primary state in the ``slow'', right-moving ($R$) chiral branch.] The leading order term driving these splittings has the form 
\begin{equation}
\label{eq:jjbarperturbation}
    \mathcal{H}^{(1)} = \sum_{n \in \mathbb{Z}}J^a_n\overline{J}^a_n.
\end{equation}
This term couples the respective Kac-Moody currents $J^a_n$ and $\overline{J}^a_n$ of the left- and right-chiral $\mathrm{SU}(3)_1$ and $\overline{\mathrm{SU}(3)_1}$ theories. Note that this term is self-conjugate and will thus have the same effect on conjugate irreps: it cannot itself be responsible for any of the splittings of conjugate irreps previously mentioned. $\mathcal{H}^{(1)}$ is thus seen to fall in the top right corner of Table \ref{table:interactionterms}. The terms of the bottom row of Table \ref{table:interactionterms}, which do split conjugate irreps, will be discussed in the next subsection.

As seen in Eqs.~\eqref{eq:htotal} and \eqref{eq:hinteraction}, $\mathcal{H}^{(1)}$ can be considered as an interaction perturbation to the left-/right- decoupled Hamiltonian $H_{\text{doubled}}$ of Eq.~\eqref{eq:htotal}.
The first order perturbative contribution to the $i$th state ($\ket{i}$) of the entanglement spectrum due to $\mathcal{H}^{(1)}$ is then
\begin{equation}
\label{eq:deltae1}
    \delta E^{(1)}_i = \lambda_1 \bra{i} \mathcal{H}^{(1)} \ket{i} = \lambda_1 \bra{i} J^a_0\overline{J}^a_0 \ket{i},
\end{equation}
for some perturbative coupling $\lambda_1$. We can write  $J^a_0\overline{J}^a_0$ as
\begin{equation}
    J^a_0\overline{J}^a_0 = \frac{(J^a_0 + \overline{J}^a_0)(J^a_0 + \overline{J}^a_0) - J^a_0J^a_0 - \overline{J}^a_0\overline{J}^a_0}{2}.
\end{equation}
Then
\begin{equation}
    \delta E^{(1)}_i = \lambda_1 \bra{i} J^a_0\overline{J}^a_0 \ket{i} =  \lambda_1\frac{\mathcal{C}^{(2)}_i - \mathcal{C}^{(2)}_{L,i} - \mathcal{C}^{(2)}_{R,i}}{2},
\end{equation}
where $\mathcal{C}^{(2)}_i$ is the quadratic Casimir invariant
of the {\it total} (diagonal) $\mathrm{SU}(3)$ representation associated to the state $\ket{i}$, while $\mathcal{C}^{(2)}_{L,i}$ and $\mathcal{C}^{(2)}_{R,i}$ are the quadratic Casimir invariants of the $\mathrm{SU}(3)$ and $\overline{\mathrm{SU}(3)}$ representations associated to the states in the left- and right-chiral CFTs that went into the tensor product that gave rise to the state $\ket{i}$ in the overall doubled theory.

Since for the tensor product of any particular pair of the left- and right-chiral multiplets $\mathcal{C}^{(2)}_{L,i}$ and $\mathcal{C}^{(2)}_{R,i}$ will be constant, the effect of $J^a_0\overline{J}^a_0$ is to arrange the irreps of the diagonal $\mathrm{SU}(3)$ within each tensor product of the doubled theory by their respective quadratic Casimirs $\mathcal{C}^{(2)}_i$, listed
\footnote{\label{footnote:irrep1}The irreps of $\mathrm{SU}(3)$ can be understood in terms of the associated Young tableaux. We can denote an irrep with a Young tableau of $r$ columns of height 1 and $s$ columns of height 2 by the notation $(r,s)$. The irrep $(r,s)$ then has dimension $d(r,s) ={\frac{1}{2}}(r+1)(s+1)(r+s+2)$ and quadratic Casimir $\mathcal{C}^{(2)}(r,s) = (r^2+s^2+3r+3s+rs)/3$. This is described in Ref.~\onlinecite{Baird1963}. (Note that Ref.~\onlinecite{Baird1963} uses instead the notation $(p,q)$ for each irrep, where $p-q =r$ and $q = s$.)\hphantom{.}
}
in Table \ref{table:quadraticcasimirs}. 
That is, the differences in the perturbative entanglement energies for states $\ket{i}$ and $\ket{j}$ at this order ($\Delta E^{(1)}_{ij}=\delta E^{(1)}_i-\delta E^{(1)}_j$) will be given by 
\begin{equation}
\label{eq:deltas}
    \Delta E^{(1)}_{ij} = \frac{\lambda_1 \Delta \mathcal{C}^{(2)}_{ij}}{2}\qquad\textrm{(where $\ket{i}$ and $\ket{j}$ come from the same tensor product),}
\end{equation}
and where $\Delta\mathcal{C}^{(2)}_{ij} = \mathcal{C}^{(2)}_i - \mathcal{C}^{(2)}_j$ is the difference in the quadratic Casimir invariants of the total $\mathrm{SU}(3)$.

\begin{figure}[H]
	\centering
	\includegraphics[scale=0.5]{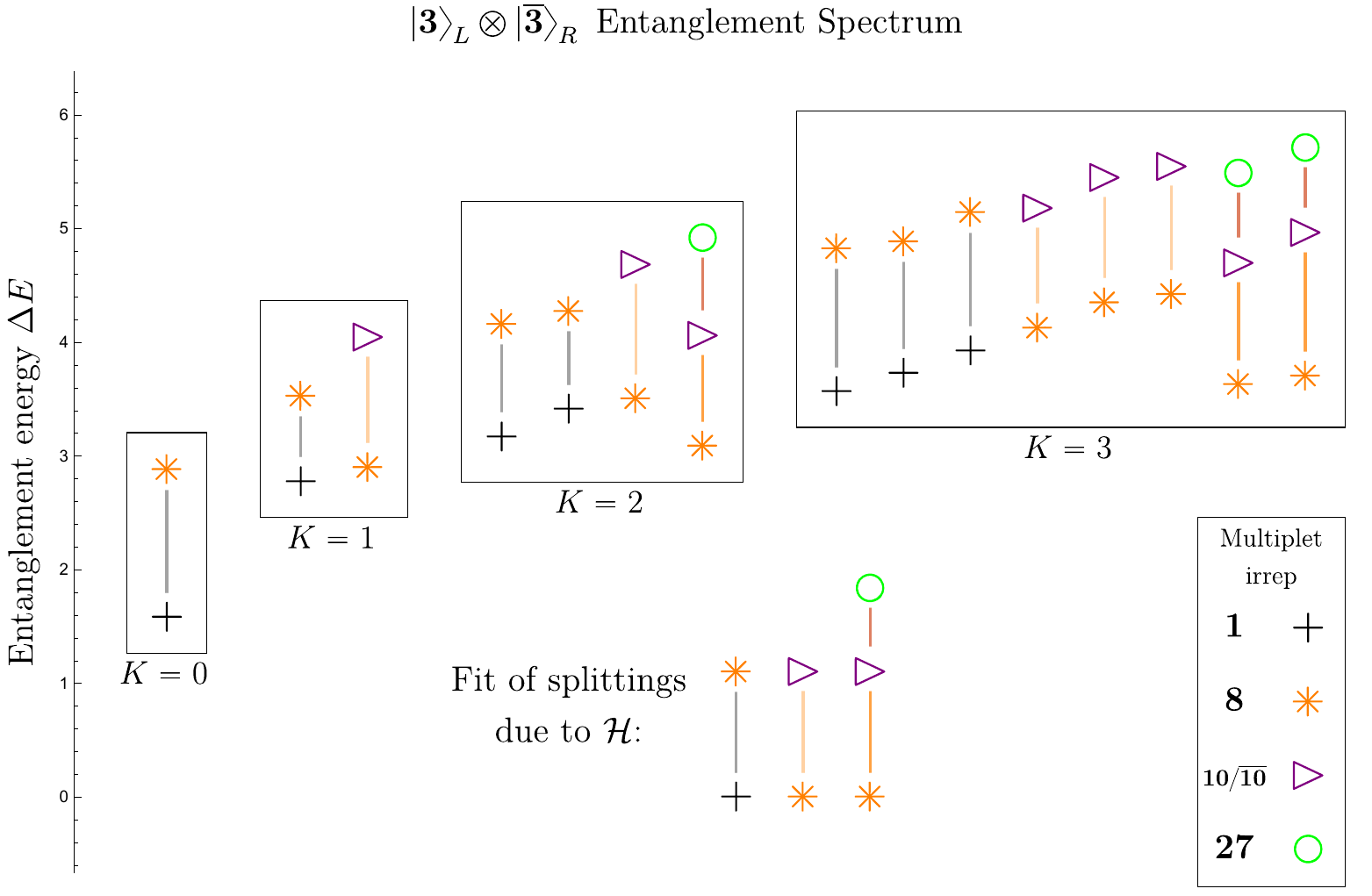}
	\caption{The entanglement spectrum of the non-chiral PEPS is shown in the $\ket{\bm{3}}_L\otimes\ket{\overline{\bm{3}}}_R$ sector. [Recall that the $L$ subscript denotes a primary state in the ``fast'', left-moving ($L$) chiral branch. An $R$ subscript denotes a primary state in the ``slow'', right-moving ($R$) chiral branch.] The splittings between multiplets are depicted by the lines between them. The fits of the different sorts of these splittings with one overall parameter multiplied by the perturbation $\mathcal{H}^{(1)}$ of Eq.~\eqref{eq:jjbarperturbation}, that is, the fit of all the differences between states within the same tensor product making use of Eq.~\eqref{eq:deltas} with a single uniform parameter $\lambda_1$ used to fit all the data, is shown at the bottom for comparison. The horizontal separation of the data points within each box at level $K$ has been artificially added in order to more clearly show overlapping data.}
	\label{fig:nonchiralessplittings02}
\end{figure}

We can directly see this in the entanglement spectrum in six of the nine sectors. The splittings are illustrated in the $\ket{\bm{3}}_L \otimes \ket{\overline{\bm{3}}}_R$ sector in Fig.~\ref{fig:nonchiralessplittings02}. (The splittings of two other sectors, $\ket{\bm{3}}_L \otimes \ket{\bm{1}}_R$ and $\ket{\bm{3}}_L \otimes \ket{\bm{3}}_R$, are additionally depicted in Appendix \ref{app:nonchiralsplittings} in Figs.~\ref{fig:nonchiralessplittings11}-\ref{fig:nonchiralessplittings20}; the remaining three sectors displaying the phenomenon are simply the conjugates of these three, for a total of six sectors where the splittings of this type are visible. 
Such splittings are {\it not} visible from the other three sectors of the nine total sectors, which have the trivial $\ket{\bm{1}}_L$ primary at the base of the high-velocity branch: $\ket{\bm{1}}_L \otimes \ket{\bm{1}}_R$, $\ket{\bm{1}}_L \otimes \ket{\bm{3}}_R$, and $\ket{\bm{1}}_L \otimes \ket{\overline{\bm{3}}}_R$. We are not able to directly see this type of splitting for these sectors in the low-energy entanglement spectrum we observe, since the states of the low-energy entanglement spectrum in these sectors will only have tensor products that simply replicate the $\mathrm{SU}(3)$ content of the low-velocity branch, due to the trivial high-velocity primary.) The linked multiplets in Fig.~\ref{fig:nonchiralessplittings02} arise from the same tensor product of multiplets in the left and right (fast and slow) chiral theories. We can observe that in Fig.~\ref{fig:nonchiralessplittings02}, the splittings of the entanglement energies $\Delta E$ of multiplets within each linked set are approximately proportional to their relative quadratic Casimirs, as denoted in the $\Delta \mathcal{C}^{(2)}$ rows of Table \ref{table:multipletbreakdownmaintext}. (A more complete version of this table containing two additional representative sectors can be found in Appendix \ref{app:nonchiralsplittings} as Table \ref{table:multipletbreakdown}.) In particular, performing a fit of Eq.~\eqref{eq:deltas} to the splittings varying the single parameter $\lambda_1$ yields an estimate of $\lambda_1 \approx 0.72$. Note that the splitting due to $\mathcal{H}^{(1)}$ does not attempt to account for the relative entanglement energies of each linked multiplet set with respect to other linked multiplet sets, which may be due to other factors including splittings present in the slow chiral theory itself. The decomposition of the respective spectral $\mathrm{SU}(3)$ multiplet content into individual tensor products, and the resulting splittings due to this perturbation, are outlined in Table \ref{table:multipletbreakdown}.

\begin{table}
    \begin{subtable}[hbt]{0.45\textwidth}
        \begin{tabular}{c|c}
    	$\mathrm{SU}(3)$ irrep & $\mathcal{C}^{(2)}$ \\
    	\hline 
    	$\bm{1}$ & 0 \\
    	$\bm{8}$ & 3 \\
    	$\bm{10}$ or $\overline{\bm{10}}$ & 6 \\
    	$\bm{27}$ & 8 
    	\end{tabular}
	\end{subtable}
	\hfill
	\begin{subtable}[hbt]{0.45\textwidth}
    	\begin{tabular}{c|c}
    	$\mathrm{SU}(3)$ irrep & $\mathcal{C}^{(2)}$ \\
    	\hline 
    	$\bm{3}$ or $\overline{\bm{3}}$ & 4/3 \\
    	$\overline{\bm{6}}$ or $\bm{6}$ & 10/3 \\
    	$\bm{15}$ or $\overline{\bm{15}}$ & 16/3 \\
    	$\bm{24}$ or $\overline{\bm{24}}$ & 25/3 \\
    	$\bm{15}'$ or $\overline{\bm{15}}'$ & 28/3 
    	\end{tabular}
	\end{subtable}
	\caption{The quadratic Casimirs $\mathcal{C}^{(2)}$ of the irreps present in the spectrum of Fig.~\ref{fig:nonchiralessplittings02} are exhibited in the left column. Those in the spectra of Figs.~\ref{fig:nonchiralessplittings11}-\ref{fig:nonchiralessplittings20} (in Appendix \ref{app:nonchiralsplittings}) are shown in the right column for completeness. Note that the quadratic Casimir is invariant under conjugation, so conjugate irreps have the same value for $\mathcal{C}^{(2)}$. (The $\bm{1}$, $\bm{8}$, and $\bm{27}$ irreps are self-conjugate.) }
\label{table:quadraticcasimirs}
\end{table}

\begin{table}
		\centering
		\begin{tabular}{c|c|c||c|x{1.7cm}|x{1.7cm}|x{1.7cm}|x{1.7cm}|x{1.7cm}}
			$(q,\phi)$ & \text{Fast ($L$)} & \text{Slow ($R$)} & \multicolumn{6}{c}{Multiplet content} \\
			\hline
			\hline
			\multirow{7}{*}{$(0,2)$} & \multirow{7}{*}{$\bm{3}$} & \multirow{7}{*}{$\overline{\bm{3}}$} & \multirow{2}{*}{Multiplets} & $\bm{3}\otimes\overline{\bm{3}}$& \multicolumn{2}{c|}{$\bm{3}\otimes\bm{6}$} & \multicolumn{2}{c}{$\bm{3}\otimes\overline{\bm{15}}$} \\
			&  &  &  & $=\bm{1}+\bm{8}$ & \multicolumn{2}{c|}{$=\bm{8}+\bm{10}$} & \multicolumn{2}{c}{$=\bm{8}+\overline{\bm{10}}+\bm{27}$}  \\
			\cline{4-9} 
			&  &  & $\Delta \mathcal{C}^{(2)}$ & 3 & \multicolumn{2}{c|}{3} & 3 & 2 \\
			\cline{4-9}
			&  &  & $K = 0$ & 1 & \multicolumn{2}{c|}{0} & \multicolumn{2}{c}{0} \\
			&  &  & $K = 1$ & 1 & \multicolumn{2}{c|}{1} & \multicolumn{2}{c}{0} \\
			&  &  & $K = 2$ & 2 & \multicolumn{2}{c|}{1} & \multicolumn{2}{c}{1} \\ 
			&  &  & $K = 3$ & 3 & \multicolumn{2}{c|}{3} & \multicolumn{2}{c}{2} \\
 			\hline
 			\hline
		\end{tabular}
\caption{The composition of the lowest four levels of the entanglement spectrum in the topological sector of the non-chiral PEPS corresponding to the $\ket{\bm{3}}_L\otimes\ket{\overline{\bm{3}}}_R$ sector of the ``doubled'' theory is shown here. The multiplet content comes from the tensor product of the fast primary $\mathrm{SU}(3)$ irrep $\ket{\bm{3}}$ with the content of the slow primary sector of descendants of $\ket{\overline{\bm{3}}}$. The value of $\Delta \mathcal{C}^{(2)}$ is given for each adjacent pair of multiplets in each tensor product: this is proportional to their splitting due to the effect of $\mathcal{H}^{(1)}$. See Eq.~\eqref{eq:deltas}. A more complete table for three representative sectors is found in the Appendix as Table \ref{table:multipletbreakdown}.}	
\label{table:multipletbreakdownmaintext}
\end{table}

\newpage

\subsection{Splittings of conjugate irreps in the non-chiral PEPS}
\label{sec:conjugate}

There will be perturbative corrections to the splittings between various irreps within each tensor product that are of
higher order than $\mathcal{H}^{(1)}$, and at such higher orders we can find perturbations that have the effect of splitting conjugate irreps in the entanglement spectrum. These correspond to the lower right corner of Table \ref{table:interactionterms}. The first such perturbations are
\begin{align}
\label{eq:jtbarperturbation}
    \mathcal{H}^{(2)} &=  \sum_{n \in \mathbb{Z}} J^a_n\overline{T}^a_n,\text{ and} \\
    \mathcal{H}^{(3)} &=  \sum_{n \in \mathbb{Z}} T^a_n\overline{J}^a_n,
\label{eq:tjbarperturbation}
\end{align}
where the $T^a_n$ are the Fourier modes of the operator $T^a(x)$ defined as\cite{Bouwknegt1993}
\begin{equation}
\label{eq:tacurrent}
    T^a(x) \propto \sum_{a,b,c =1}^8 d_{abc} (J^b J^c)(x).
\end{equation}
Note that this implies [see Eq.~\eqref{eq:wcurrent}] that
\begin{equation}
    W(x) \propto \sum_{a=1}^8 (J^a T^a)(x).
\end{equation}

Therefore, while $\mathcal{H}^{(1)}$ reveals the quadratic Casimir invariant of the global $\mathrm{SU}(3)$ irrep, it turns out that $\mathcal{H}^{(2)}$ and $\mathcal{H}^{(3)}$ are related to the cubic Casimir invariant. And unlike $\mathcal{H}^{(1)}$, $\mathcal{H}^{(2)}$ and $\mathcal{H}^{(3)}$ are not necessarily invariant under conjugation and may thus have different effects in conjugate sectors. 
Note also that there may additionally be other perturbations, of higher order than $\mathcal{H}^{(1)}$, $\mathcal{H}^{(2)}$, and $\mathcal{H}^{(3)}$, which contribute to the splittings of conjugate states that arise from the same tensor product. These can be found by combining the operators of $\mathcal{H}^{(1)}$, $\mathcal{H}^{(2)}$, and $\mathcal{H}^{(3)}$ with chiral operators, on either side, that are $\mathrm{SU}(3)$ singlets. (Two such terms will constitute $\mathcal{H}^{(4)}$ and $\mathcal{H}^{(5)}$, as will be discussed later.)

\begin{figure}[H]
	\centering
	\includegraphics[angle=90,origin=c,scale=0.6]{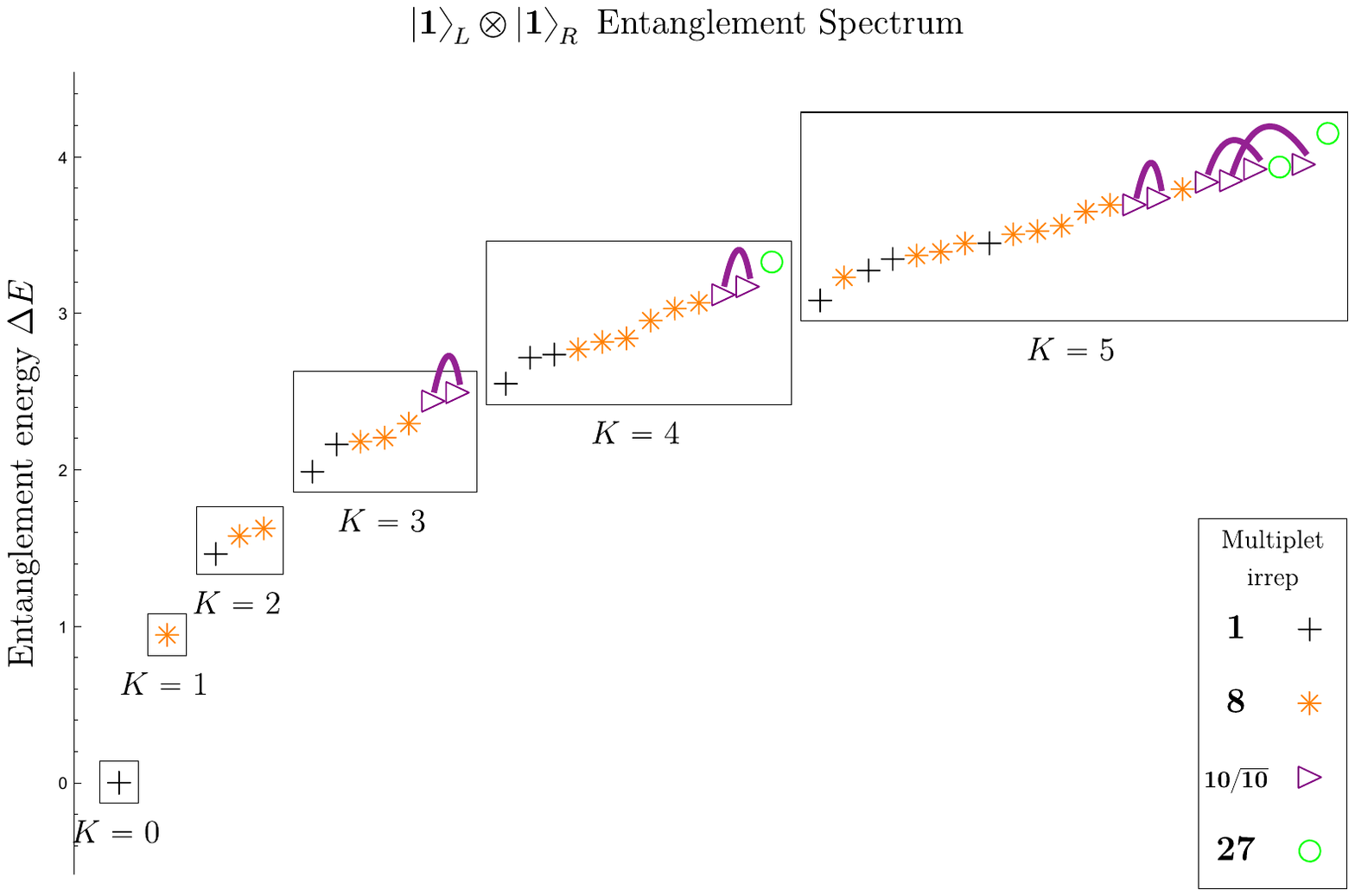}
	\caption{The entanglement spectrum of the non-chiral PEPS is shown in the $\ket{\bm{1}}_L\otimes\ket{\bm{1}}_R$ sector. [Recall that the $L$ subscript denotes a primary state in the ``fast'', left-moving ($L$) chiral branch. An $R$ subscript denotes a primary state in the ``slow'', right-moving ($R$) chiral branch.] The slight breaking of the degeneracy between $\bm{10}$ and $\overline{\bm{10}}$ multiplets (indicated by the purple braces) is visible. This may be compared to the degeneracy of $\bm{10}$ and $\overline{\bm{10}}$ multiplets observed in the $Q=0$ chiral entanglement spectrum (Fig.~\ref{fig:chirales0fit}). The horizontal separation of the data points within each box at level $K$ has been artificially added in order to more clearly show overlapping data.}
	\label{fig:nonchiralessplittings00}
\end{figure}

\begin{figure}[H]
	\centering
	\includegraphics[scale=0.5]{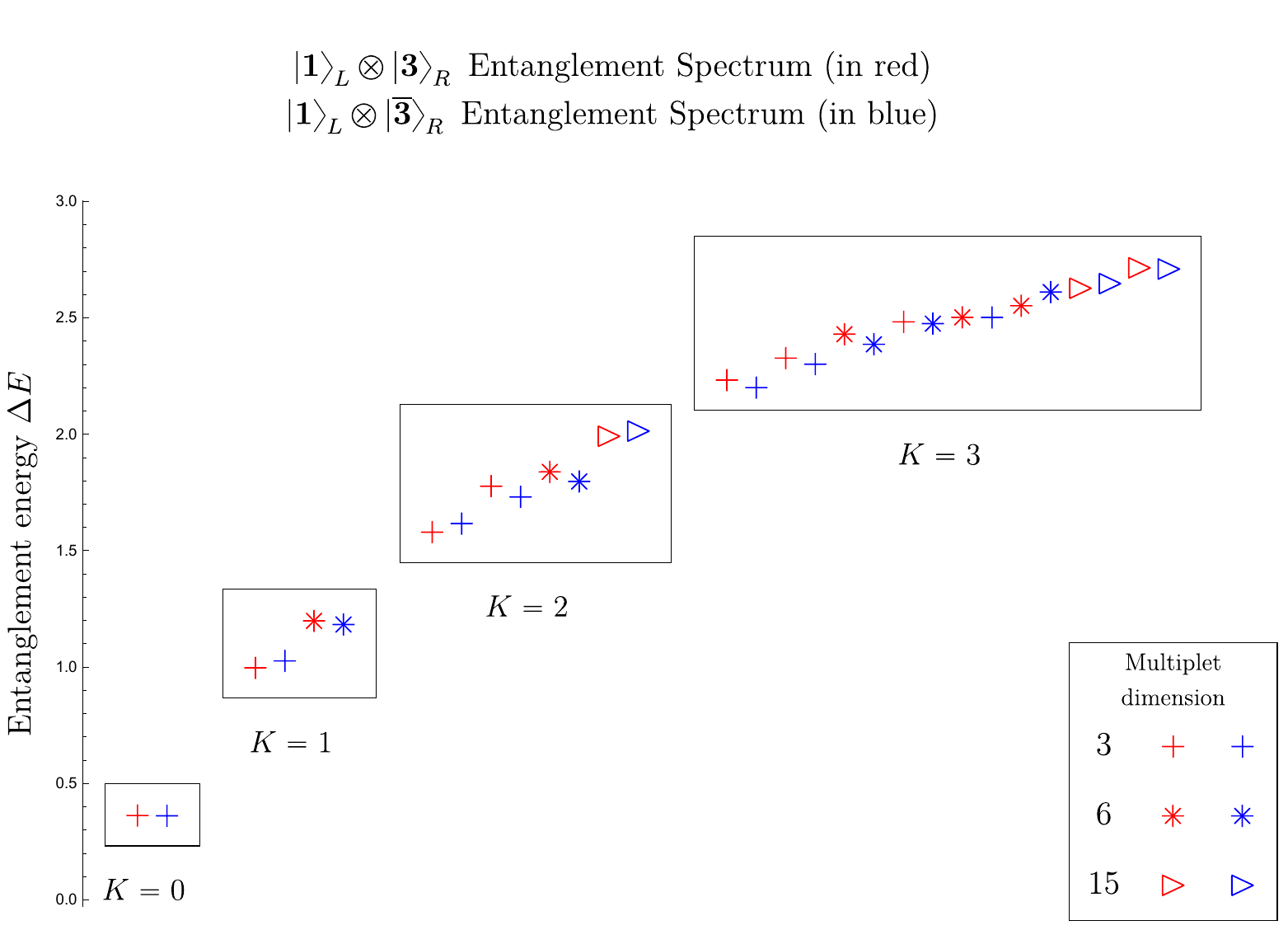}
	\caption{The entanglement spectrum of the non-chiral PEPS is shown in the $\ket{\bm{1}}_L\otimes\ket{\bm{3}}_R$ and $\ket{\bm{1}}_L\otimes\ket{\overline{\bm{3}}}_R$ sectors. [Recall that the $L$ subscript denotes a primary state in the ``fast'', left-moving ($L$) chiral branch. An $R$ subscript denotes a primary state in the ``slow'', right-moving ($R$) chiral branch.] The multiplets of these sectors are depicted in red and blue, respectively. The slight breaking of the degeneracy between conjugate irreps in these two conjugate sectors is visible. This may be compared to the $Q = \pm 1$ chiral spectra of the chiral PEPS (e.g., Fig.~\ref{fig:chirales1fit}). In that case, the $Q = 1$ and $Q = -1$ sectors consisted of closely degenerate states. The horizontal separation of the data points within each box at level $K$ has been artificially added in order to more clearly show overlapping data. (The red and blue color coding in this figure has no relationship with the red and blue color coding used in Fig.~\ref{fig:boundarycylindergraphic}.)}
	\label{fig:nonchiralessplittings1221}
\end{figure}

And indeed, in the non-chiral PEPS, we do in fact observe breaking of the degeneracy of conjugate states in the entanglement spectrum. This can be seen, for instance, with the slight (but clearly visible) breaking of the degeneracy between the $\bm{10}$ and $\overline{\bm{10}}$ states in the in the topological sector of the non-chiral PEPS corresponding to the $\ket{\bm{1}}_L \otimes \ket{\bm{1}}_R$ sector of the ``doubled'' theory, depicted in Fig.~\ref{fig:nonchiralessplittings00} (compare the exact degeneracy between the $\bm{10}$ and $\overline{\bm{10}}$ states necessarily found in the corresponding {\it chiral} $\mathrm{SU}(3)_1$ PEPS entanglement spectrum of Fig.~\ref{fig:chirales0fit}, as discussed in Sec.~\ref{sec:results}), or of the degeneracy between the states in the entanglement spectra of the other eight sectors accessible in that PEPS and the states of the entanglement spectra of those sectors' respective conjugate partners. The $\ket{\bm{1}}_L \otimes \ket{\bm{3}}_R$ and $\ket{\bm{1}}_L \otimes \ket{\overline{\bm{3}}}_R$ sectors of the non-chiral PEPS, which mimic chiral spectra, are depicted next to each other in Fig.~\ref{fig:nonchiralessplittings1221} so that the breaking of their degeneracy is apparent. 
On the other hand, in the chiral $\mathrm{SU}(3)_1$ PEPS data of Ref.~\onlinecite{Chen2020}, the spectra of the corresponding $\ket{\bm{3}}$ and $\ket{\overline{\bm{3}}}$ primary sectors are degenerate.
\footnote{This data is depicted in Fig.~\ref{fig:chirales1fit}, for instance, where the degenerate levels are sufficiently close that only one was necessary to be drawn.} 
Table \ref{table:chiraldegeneracycomparison} shows, for comparison, the numerical entanglement energy data of the splittings of conjugate states in the comparable sectors of both the non-chiral PEPS (left column) and chiral PEPS (right column). 
While in the numerical data, the degeneracy in the chiral case is not found to be exactly zero (presumably due to the effects of the numerical calculation), it is seen to be orders of magnitude smaller than that in the non-chiral case.

\begin{table}[]
    \centering
    \begin{tabular}{c|c|c|c|c}
        Sector & $K$ & Irrep & Non-chiral ES & Chiral ES\\
        \hline
        \multirow{5}{*}{$\ket{\bm{1}}$} & 3 & \multirow{5}{*}{$\bm{10}$/$\overline{\bm{10}}$} & 0.0543147 & 0.0000192286 \\
        & 4 & & 0.0520877 & 0.0000125578 \\
        & 5 & & 0.042824 & $2.082225716293351\times 10^{-6}$ \\
        & 5 & & 0.0831486 & 0.0000140543 \\
        & 5 & & 0.103837 & 0.000238927 \\
        \hline
        \multirow{7}{*}{$\ket{\bm{3}}$/$\ket{\overline{\bm{3}}}$} & 0 & $\bm{3}$/$\overline{\bm{3}}$ & -0.00148651 & 0. \\
        & 1 & $\bm{3}/\overline{\bm{3}}$ & 0.0300456 & 0.0000445773 \\
        & 1 & $\overline{\bm{6}}$/$\bm{6}$ & -0.0154502 & 0.0000264267 \\
        & 2 & \multirow{2}{*}{$\bm{3}/\overline{\bm{3}}$} & 0.0367921 & 0.0000423184 \\
        & 2 & & -0.0460727 & 0.00016052 \\
        & 2 & $\overline{\bm{6}}$/$\bm{6}$ & -0.0412642 & 0.000040769 \\
        & 2 & $\bm{15}/\overline{\bm{15}}$ & 0.0221733 & 0.000103927 \\
    \end{tabular}
    
    \caption{The differences in entanglement energy between entanglement spectrum levels with conjugate $\mathrm{SU}(3)$ irreps are shown for the entanglement spectra of both the non-chiral and chiral PEPS in the sectors where the comparison is reasonable---that is, the sectors where both non-chiral and chiral PEPS ES exhibit the chiral ($\mathrm{SU}(3)_1$) Li-Haldane state counting. The ``Sector'' column lists the primary state of the sector in the chiral ES, which is the primary state of the low-velocity sector for the non-chiral ES. 
    (In the relevant sectors of the non-chiral ES here, the high-velocity sector will simply be the trivial primary state sector rooted at $\ket{\bm{1}}$.) 
    Note that the entanglement energy differences between conjugates are orders of magnitude smaller for the chiral ES.}
    \label{table:chiraldegeneracycomparison}
\end{table}

The breaking of this degeneracy can occur because, in the non-chiral setting like that in Refs.~\onlinecite{Kurecic2019} and \onlinecite{ArildsenSchuchLudwig2022}, it is possible for the entanglement spectrum to possess contributions from terms in the entanglement Hamiltonian that break the conjugation symmetry of the entanglement spectrum. One can consider the effect of compound perturbations that feature operators from both the left- and right-moving theories. These perturbations, by contrast, are not possible in the \emph{chiral} entanglement spectrum, where the integrals of operators of only one chirality, only left- or only right-moving, are available, as detailed in Table \ref{table:operators}.

The perturbations $\mathcal{H}^{(2)}$ and $\mathcal{H}^{(3)}$ shown in Eqs.~\eqref{eq:jtbarperturbation}--\eqref{eq:tjbarperturbation}, which do distinguish between conjugate states in the case where splittings are between irreps that come from the tensor product of the same $\mathrm{SU}(3)$ irreps in the chiral left- and right-moving theories, are two such interaction terms.

\begin{figure}[H]
	\centering
	\includegraphics[scale=0.6]{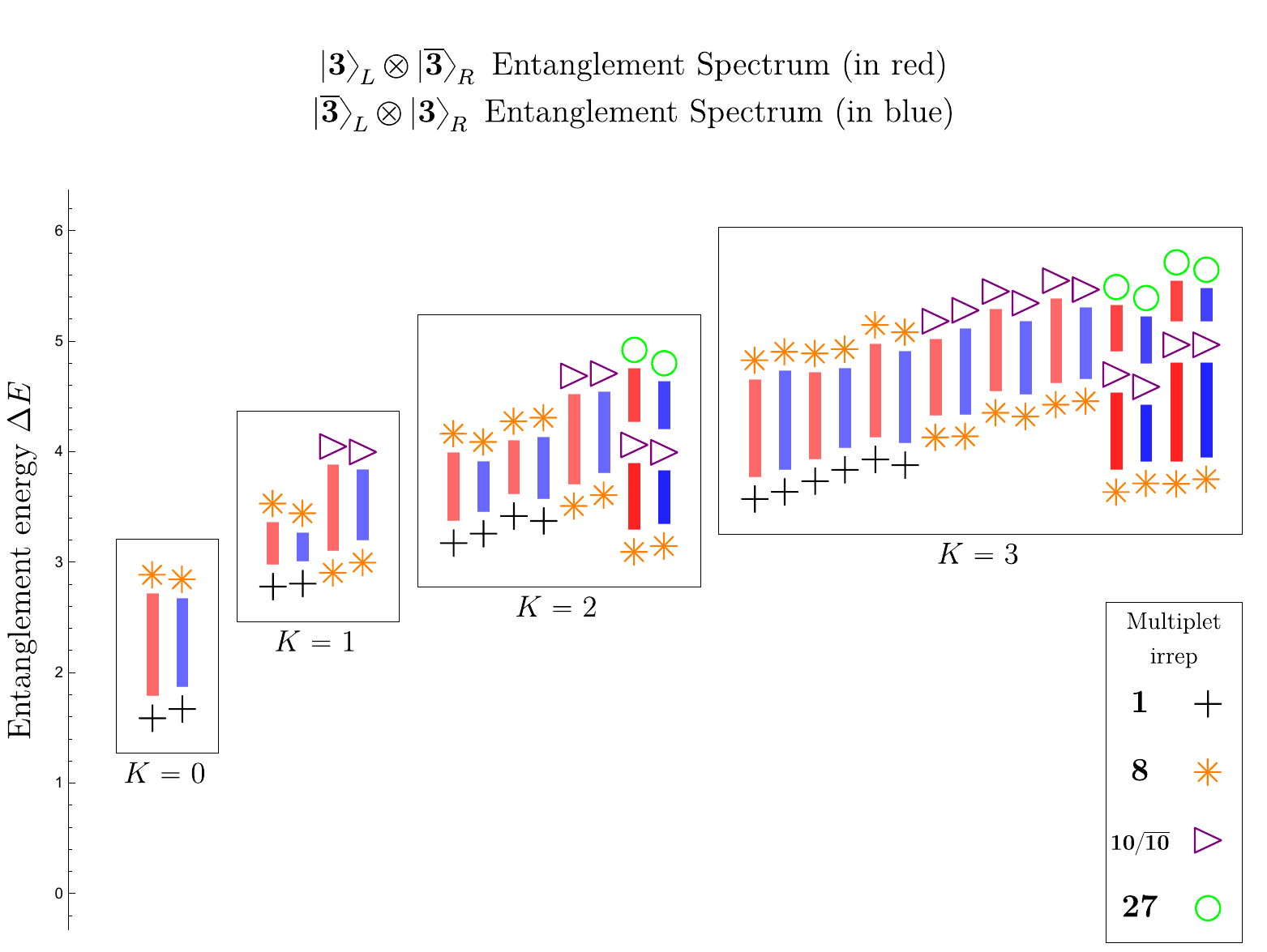}
	\caption{The entanglement spectrum of the non-chiral PEPS is shown in both the $\ket{\bm{3}}_L\otimes\ket{\overline{\bm{3}}}_R$ and $\ket{\overline{\bm{3}}}_L\otimes\ket{\bm{3}}_R$ sectors, which are conjugate. [Recall that the $L$ subscript denotes a primary state in the ``fast'', left-moving ($L$) chiral branch. An $R$ subscript denotes a primary state in the ``slow'', right-moving ($R$) chiral branch.] The splittings between multiplets in a given tensor product are depicted by the bars between them: these are red for the spectrum in the $\ket{\bm{3}}_L\otimes\ket{\overline{\bm{3}}}_R$ sector and blue for the spectrum in the $\ket{\overline{\bm{3}}}_L\otimes\ket{\bm{3}}_R$ sector. The adjacent pairs of split multiplets may properly be compared. The horizontal separation of the data points within each box at level $K$ has been artificially added in order to more clearly show overlapping data. (The red and blue color coding in this figure, like that in Fig.~\ref{fig:nonchiralessplittings1221}, has no relationship with the red and blue color coding used in Fig.~\ref{fig:boundarycylindergraphic}.)}
	\label{fig:nonchiralesconjsplittings02}
\end{figure}

Another two such terms, contributing at a higher order, are
\begin{align}
     \mathcal{H}^{(4)} &= \sum_{n \in \mathbb{Z}}J^a_{n}\left(\overline{J}^a \overline{W}\right)_n,\text{ and} \\
     \mathcal{H}^{(5)} &= \sum_{n \in \mathbb{Z}}\left(J^a W\right)_{n}\overline{J}^a_n,
\end{align} which act both to split conjugate multiplets, through the action of the $W(x)$ or $\overline{W}(x)$ modes, and to split multiplets associated with the same tensor product of the chiral theories, through a similar mechanism to that of $\mathcal{H}^{(1)}$. We can see both effects in Fig.~\ref{fig:nonchiralesconjsplittings02}, where there are slight differences between the two conjugate sectors (in red, for the $\ket{\bm{3}}_L\otimes\ket{\overline{\bm{3}}}_R$ sector, and in blue, for the $\ket{\overline{\bm{3}}}_L\otimes\ket{\bm{3}}_R$ sector) in the splittings between the various $\mathrm{SU}(3)$ irreps that arise from the same tensor product of irreps in the left- and right-chiral theories. We thus see that unlike in the chiral case, it is possible to have a breaking of the degeneracy of the conjugate irreps in the non-chiral case, due to these compound perturbations. 

But the breaking of the conjugation symmetry occurs even in sectors where the left-moving (``high-velocity/fast'') primary state is a singlet (``trivial''), and thus the low-lying entanglement spectrum consists of just the $\mathrm{SU}(3)$ multiplet content of the chiral right-moving (``low-velocity/slow'')CFT. Thus, in order to understand possible terms in the entanglement Hamiltonian responsible for conjugacy splitting, we must turn our attention to another class of interaction terms: those which actually preserve the $\mathrm{SU}(3) \times \overline{\mathrm{SU}(3)}$ symmetry of both the left- and right-moving theories. These are those in the left column of Table \ref{table:interactionterms}. 
They consist of couplings of modes of the chiral operators of Table \ref{table:operators} and their antichiral counterparts. The pairing of $T(x)$ and $\overline{T}(x)$ is the coupling of this kind with lowest dimension, but such a term will not split conjugate irreps. It is thus represented in the top left corner of Table \ref{table:interactionterms}. In order to split conjugate irreps, we must take advantage of $W(x)$, the integral of which (as previously noted in Sec.~\ref{sec:results}) is known to do so. We can then consider the effect of terms like
\begin{align}
    \mathcal{H}^{(6)} &= \sum_{n \in \mathbb{Z}} \left(L_n-\frac{c}{24}\delta_{n,0}\right)\overline{W}_n\text{ and} \\
    \mathcal{H}^{(7)} &= \sum_{n \in \mathbb{Z}} W_n\left(\overline{L}_n-\frac{c}{24}\delta_{n,0}\right).
\end{align}
Non-chiral terms like these will indeed be able to split conjugate irreps due to their $W_n$ and $\overline{W}_n$ modes. They are thus represented in the lower left corner of Table \ref{table:interactionterms}.

One can in principle use all of the $\mathcal{H}^{(i)}$ above to gain some understanding of the splittings of the various multiplets by calculating the associated first order perturbations to the $j$th state of the entanglement spectrum ($\ket{j}$):
\begin{equation}
\label{eq:deltaei}
    \delta E^{(i)}_j = \lambda_i \bra{j} \mathcal{H}^{(i)} \ket{j},
\end{equation}
(This is analogous to the calculation of the $\delta E_i^{(1)}$ in Eq.~\eqref{eq:deltae1} from $\mathcal{H}^{(1)}$.) 

\newpage

\newpage

\section{Conclusions}
\label{sec:conclusion}

In the present work, building on our understanding from previous work in Ref.~\onlinecite{ArildsenLudwig2022} on a Generalized Gibbs Ensemble approach to the entanglement spectrum, we have been able to comprehend the structure of the splittings in the low-lying ES of levels degenerate in momentum, of Abelian chiral $\mathrm{SU}(3)$ spin liquid states in (2+1)-dimensions at finite size, purely from the point of view of the corresponding chiral $\mathrm{SU}(3)$-level-one (1+1)-dimensional CFT.
This approach lets us view the entanglement Hamiltonian as a particular linear combination of conserved quantities from that CFT, which reflects the specific wavefunction within the topological phase in its ES. We determine the particular coefficients for the leading terms of this linear combination by fitting the entanglement spectrum data of the Abelian chiral $\mathrm{SU}(3)$ spin liquid PEPS wavefunction of Chen {\it et al.}~from Ref.~\onlinecite{Chen2020}, thereby ``mapping out'' features of this wavefunction in its ES. 
This fine-grained approach enables us to understand the entanglement spectra through the lens of both the global $\mathrm{SU}(3)$ and the discrete symmetry (in particular, preservation of neither time-reversal $\mathcal{T}$ nor reflection $\mathcal{R}$ alone, but preservation of their product $\mathcal{RT}$) present in the states we consider in this paper.

Specifically, we are able to use the splitting of conjugate $\mathrm{SU}(3)$ irreps in the low-lying entanglement spectrum to diagnose whether the underlying topological state is chiral or non-chiral, as we show that a splitting of conjugate $\mathrm{SU}(3)$ irreps is not possible for a chiral $\mathrm{SU}(3)$-level-one entanglement spectrum with these discrete symmetries. We demonstrate this diagnostic by comparing the entanglement spectrum from the data of the Abelian {\it chiral} $\mathrm{SU}(3)$ spin liquid PEPS of Ref.~\onlinecite{Chen2020} to that of the {\it non-chiral} Abelian $\mathrm{SU}(3)$ spin liquid PEPS with the same symmetry properties under time-reversal ${\cal T}$, reflection ${\cal R}$, and its product $\mathcal{RT}$, found in Ref.~\onlinecite{Kurecic2019}: We find that, indeed, the former, chiral spectrum possesses no splittings of conjugate irreps, while the latter, non-chiral spectrum does possess such splittings. We also explain that coupling between the left- and right-moving chiral branches of the ES of the non-chiral state, which is present in the non-chiral entanglement Hamiltonian, can indeed give rise to such splittings.

This method of analyzing conjugate splittings to assess chirality has the advantage of being straightforward to read off from the entanglement spectrum. Other methods of directly establishing a non-vanishing chiral central charge exist, such as the modular commutator\cite{Kim2022chiral,Kim2022modular,Zou2022}, but the simplicity of the conjugate splitting approach makes it numerically and practically very accessible, as the entanglement spectrum is already known to be calculable numerically for relatively complex wavefunctions such as a PEPS (as in the data considered in this paper) at numerically attainable system sizes that contain detailed information about the underlying topological state. In future, we hope that this method can be used to conduct studies of the chirality of families of $\mathrm{SU}(3)$ PEPS going beyond the particular chiral $\mathrm{SU}(3)$ spin liquid PEPS of Ref.~\onlinecite{Chen2020} explored here. 
Further, the phenomenon of the absence of conjugate splittings should be a more general feature of chiral $\mathrm{SU}(N)$ spin liquid PEPS for $N > 2$. Extension to chiral $\mathrm{SU}(4)$ spin liquid PEPS is already in progress, and we hope to explore this in other systems as well, subject to available numerical entanglement spectrum data.
\footnote{Such data is now available for higher-$N$ Abelian spin liquid PEPS: see, e.g., Ref.~\onlinecite{Chen2021}.}

\acknowledgments

M.J.A., N.S., and A.W.W.L.\ would like to thank the Erwin Schr\"{o}dinger International Institute for Mathematics and Physics (ESI) in Vienna, Austria, for hospitality and useful discussion during the ``Tensor Networks: Mathematical Structures and Novel Algorithms'' thematic program from August 29 to October 21, 2022, and in particular during the ``Mathematical structure of Tensor Networks'' workshop from October 3--7, 2022.
M.J.A.\ acknowledges financial support from the PNRR MUR project PE0000023-NQSTI.
J.-Y.C.\ was supported by National Natural Science Foundation of China (NSFC) (Grant No.~12304186). 
N.S.\ acknowledges support by the European Union’s Horizon 2020 program through
the ERC-CoG SEQUAM (Grant No.\ 863476).
Part of the numerical calculations were carried out on the Vienna Scientific Cluster (VSC).

\newpage

\appendix
\section{Calculation of the Conserved Quantities}
\label{app:calculation}

To calculate the conserved quantities $H^{(i)} \propto \int \Phi_i(x) dx$ of the GGE of Eq.~\eqref{eq:realrhol}, we write down the integrals of the currents in Table \ref{table:operators} in terms of the Fourier modes of the corresponding currents and act on basis states of the chiral $\mathrm{SU}(3)_1$ WZW Hilbert space, also written in terms of modes. (The diagonalized linear combination of the $H^{(i)}$ in Eq.~\eqref{eq:lincomb} yields the splittings in the entanglement spectrum.) 
Explicit expressions for the $H^{(i)}$ are listed in Table \ref{table:modereps}. One approach to this procedure would be to expand the $L_n$ [modes of $T(x)$] and $W_n$ [modes of $W(x)$] in terms of $J^a_n$ [modes of the Kac-Moody currents $J^a(x)$] based on the Sugawara form Eq.~\eqref{eq:sugawara} of $T(x)$ and Eq.~\eqref{eq:wcurrent} for $W(x)$, and then to act with these on basis states of the form in Eq.~\eqref{eq:jbasis}. However, for the chiral $\mathrm{SU}(3)_1$ WZW theory, there is a simpler approach, taking advantage of the $\mathrm{SU}(3)$ symmetry and Abelian characteristics of the theory. In essence, the chiral $\mathrm{SU}(3)_1$ WZW theory (with central charge $c = 2$) can be reformulated as a theory of two free bosons. 

As described in Sec.~\ref{sec:wzw}, the states of the Hilbert space can be arranged in $\mathrm{SU}(3)$ representations, or multiplets. Our locally conserved quantities will all take the same value within each of these multiplets. States within each multiplet can be described by a pair of $\mathrm{U}(1)$ quantum numbers, which effectively define a coordinate lattice in two dimensions.\cite{Hall2016} We can refer to as ``central'' states the states in each descendant multiplet which share $\mathrm{U}(1)$ quantum numbers with the highest weight states of the primary state $\ket{\bm{1}}$, $\ket{\bm{3}}$, or $\ket{\overline{\bm{3}}}$, depending on the sector. As it turns out, we can in fact write a complete basis for these ``central'' states of the Hilbert space using only the $J^7_n$ and $J^8_n$ modes of the $\mathrm{SU}(3)$ current (generating the Cartan subalgebra) to build up from the three primary states, rather than all 8 types of modes $J^a_n$. Furthermore, rather than Eqs.~\eqref{eq:sugawara} and \eqref{eq:wcurrent}, we can instead write\cite{Bouwknegt1993} 
\begin{align}
    T(x) &= \frac{1}{2}\left(J^7 J^7 + J^8 J^8\right)(x) \\
    W(x) &= -\frac{1}{6}\left(3 J^7 J^7 J^8 - J^8 J^8 J^8\right)(x).
\end{align}
Thus we can write the $L_n$ and $W_n$ modes solely in terms of the $J^7_n$ and $J^8_n$ as well. This allows us to write the $H^{(i)}$ solely in terms of these modes (using the expressions in Table \ref{table:modereps}). We can then use the commutation relations of the $J^7_n$ and $J^8_n$ modes to compute the matrix elements of the $H^{(i)}$ in the aforementioned basis of $J^7_n$ and $J^8_n$ modes acting on the primary states.

\begin{table}[hbt]
	\centering
	\begin{tabular}{c | c | c | c}
		$i$ & $\Delta_i$ & $\Phi_i(x)$ & $\tilde{H}^{(i)} \propto \int \Phi_i(x) dx$ \\
		\hline
		\hline
		1 & 2 & $T(x)$ & $L_0 - \frac{c}{24}$\\
		\hline
		\rowcolor{Gray}
		2 & 3 & $W(x)$ & $W_0$ \\
		\hline
		3 & 4 & $(TT)(x)$ & $2\sum_{n>0} L_{-n} L_n + L_0^2 - \frac{c}{12}L_0 + \frac{c^2}{576} $  \\
		\hline
		\rowcolor{Gray}
		4 & 5 & $(TW)(x)$ & $\sum_{n>0} (L_{-n}W_n + W_{-n}L_n) + L_0 W_0 - \frac{c}{24}W_0 $ \\
		\hline
		\multirow{2}{*}{5} & \multirow{2}{*}{6} & \multirow{2}{*}{$(T(TT))(x)$} & $\sum_{n_1+n_2+n_3=0}:L_{n_1} L_{n_2} L_{n_3}: + \frac{3}{2}\sum_{n>0} n^2 L_{-n} L_n + \frac{3}{2}\sum_{n>0} L_{1-2n} L_{2n-1}$ \\
		& & & $-\frac{c}{4}\sum_{n>0} L_{-n}L_n - \frac{c}{8}L_0^2 + \frac{c^2}{192}L_0 -\frac{c^3}{13824}$ \\
		\hline
	    6 & 6 & $(\partial T \partial T)(x)$ & $2 \sum_{n>0} n^2 L_{-n} L_n$ \\
		\hline 
		\rowcolor{Gray}
		7 & 6 & $(T\partial W)(x)$ & $i \sum_{n > 0}n (- L_{-n}W_n + W_{-n}L_n)$ \\
		\hline
		8 & 6 & $(WW)(x)$ & $2\sum_{n>0} W_{-n} W_n + W_0^2$  \\
		\hline
	\end{tabular}
	\caption{
	Enumeration of the size-independent parts $\tilde{H}^{(i)}$ of the corresponding $\mathrm{SU}(3)$-invariant locally conserved quantities $H^{(i)}$ written in terms of the Fourier modes $L_n$ and $W_n$ of the energy-momentum tensor $T(x)$ and the currents $W(x)$, respectively, up to conformal dimension $\Delta \leq 6$. The grayed-out rows are excluded from the entanglement spectrum by consideration of discrete symmetries. The index\footnote{The index $i$ should not be confused with the imaginary number $i$ found in some entries.} $i$ denotes the quantity the parameter $\beta_i$ [in, e.g., Eq.~\eqref{eq:lincomb}] refers to. $\Delta_i$ is the conformal dimension of the operator $\Phi_i(x)$ integrated to yield the conserved quantity $H^{(i)}$. The symbols $::$ in the $i = 5$ row mean we normal order in increasing order of the subscripts $n_1,n_2,n_3$. For the $\mathrm{SU}(3)_1$ case considered here, the central charge $c = 2$.}
	\label{table:modereps}
\end{table} 

Performing these diagonalizations, we can see that of the conserved quantities in Table \ref{table:modereps}, only $\tilde{H}^{(2)}$ and $\tilde{H}^{(4)}$ have the property that they have eigenvalues of opposite sign on conjugate irreps of $\mathrm{SU}(3)$, while the others have the same eigenvalue on conjugate irreps. We can understand this as follows. The transformation that takes states in the chiral $\mathrm{SU}(3)_1$ Hilbert space to states in a conjugate irrep is given by\footnote{This transformation essentially has the effect of turning the ``raising operators'' within each irrep into ``lowering operators'' and vice-versa. The effect on the $\mathrm{SU}(3)$ algebra of the zero-modes $J^a_0$ amounts to an outer automorphism of that algebra, which negates the cubic Casimir.\cite{Biedenharn1963}\hphantom{.}}
\begin{align}
    J^a(x) \mapsto -J^a(x), & & \text{and} & & i \mapsto -i.
\end{align}
Under this transformation, modes of $W(x)$, which have an odd number of factors of $J^a(x)$, will get an overall minus sign, while modes of $T(x)$ will remain invariant. Thus, of the conserved quantities in Table \ref{table:modereps}, $\tilde{H}^{(2)}$ and $\tilde{H}^{(4)}$ will have opposite eigenvalues on conjugate pairs of irreps. $\tilde{H}^{(7)}$, however, will remain invariant due to the factor of $i$ out front. That $i$ comes from the single derivative in $(T\partial W)(x)$, and so we can state a general principle: 
{\it the conserved quantities that will split conjugate pairs of irreps are exactly the integrals of those $\Phi_i(x)$ for which the sum of the number of derivatives and the number of factors of $W(x)$ is odd. In particular, as derivatives and factors of $W(x)$ both have odd contributions to the conformal dimension of the $\Phi_i(x)$, these conserved quantities will be exactly those that are the integrals of operators of odd conformal dimension. And as it turns out, these are exactly the criteria for exclusion of these $\Phi_i(x)$ from the GGE we consider due to the restrictions imposed by $\mathcal{RT}$ symmetry, as discussed in the next Appendix.}

\section{Results Regarding the Exclusion of Conserved Quantities due to Discrete Symmetries}
\label{app:discrete}

The chiral $\mathrm{SU}(3)_1$ current $J^a(x)$ is odd under the chiral $\mathcal{RT}$ transformation. We can see this by writing $J^a(x)$ in its full time-dependent form $J^a(t,x)$ and using the fact that the operator has scaling dimension $\Delta = 1$:
\begin{align}
   \label{eq:RTJop}
   (\mathcal{RT} )J^a(t,x) (\mathcal{RT})^{-1} &= J^a(-t, -x) = (-1)^1 J^a(t,x) = -J^a(t,x).
\end{align}
Thus, in particular, at fixed time,
\begin{equation}
(\mathcal{RT})J^a(x)(\mathcal{RT})^{-1} = -J^a(x).
\end{equation}
This passes through to the mode expansion Eq.~\eqref{eq:modeexpansion}, so we have
\begin{equation}
    (\mathcal{RT})J^a_{n}(\mathcal{RT})^{-1} = -J^a_{n}.
\end{equation}
Since Eq.~\eqref{eq:sugawara} implies that there will be two modes $J^a_n$ in each term of the expression for the Virasoro modes $L_n$, we will also have 
\begin{equation}
    (\mathcal{RT})L_{n}(\mathcal{RT})^{-1} = L_{n}.
\end{equation}
[This could also be derived directly from the scaling dimension of $T(x)$ in a manner analogous to Eq.~\eqref{eq:RTJop}.]
By the same logic applied to Eq.~\eqref{eq:wcurrent}, given that there will be three $J^a_n$ in each term of the expression for the modes $W_n$,
we get that
\begin{equation}
    (\mathcal{RT})W_{n}(\mathcal{RT})^{-1} = -W_{n}.
\end{equation}
Hence, terms with odd numbers of modes $W_n$ will be odd under the $\mathcal{RT}$ transformation. This means that, of the conserved quantities in Table \ref{table:modereps}, both $\tilde{H}^{(2)}$ (the integral of $W(x)$) and $\tilde{H}^{(4)}$ (the integral of $(TW)(x)$) will be odd under $\mathcal{RT}$, and therefore excluded from consideration in the chiral $\mathrm{SU}(3)_1$ case. $\tilde{H}^{(7)}$ has an odd number of modes $W_n$ but is in fact even under $\mathcal{RT}$ due to the leading factor of $i$, which is inverted under time-reversal, leading to overall invariance, so this consideration alone is insufficient to explain its empirically observed exclusion from the ensemble of allowed conserved quantities.

\section{Further Fits to the Entanglement Spectrum of the Chiral $\mathrm{SU}(3)$ Spin Liquid PEPS}
\label{sec:furtherfits}

\begin{figure}[H]
	\centering
	\subfloat[]{\label{fig:furtherchirales1centerfit} \includegraphics[scale=0.4]{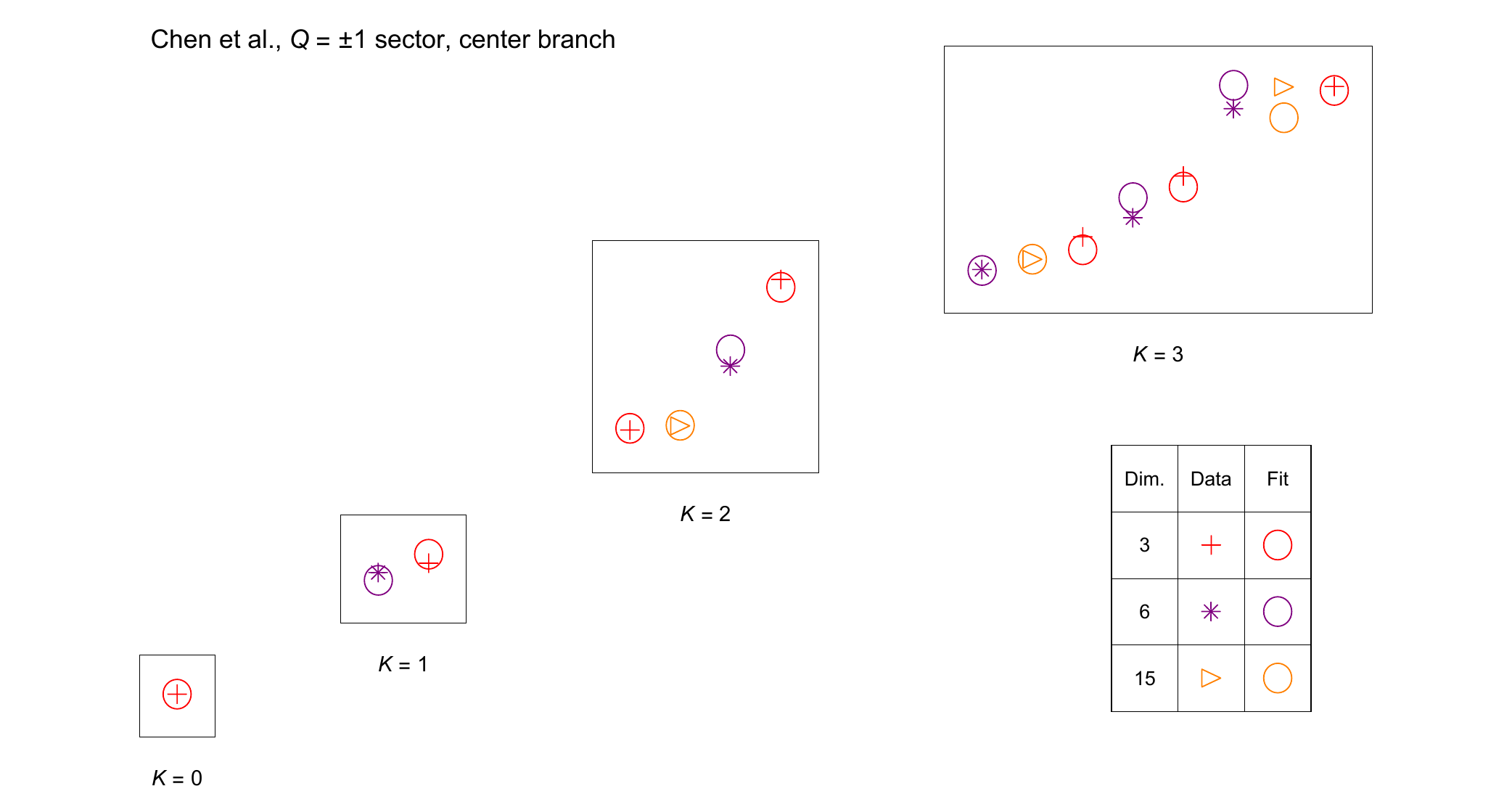}
	} \\
	\subfloat[]{\label{fig:furtherchirales1rightfit}
	\includegraphics[scale=0.4]{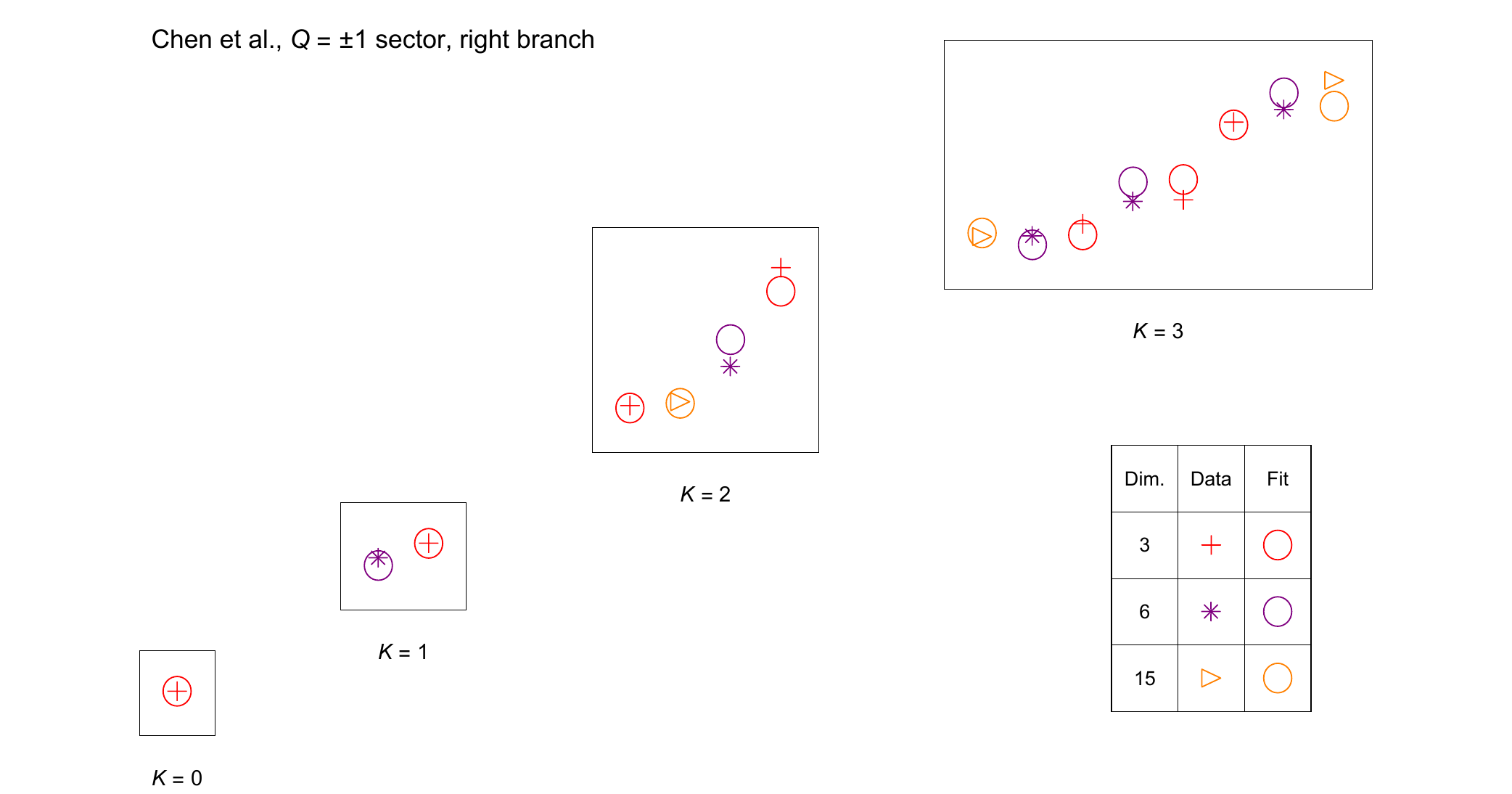}
	}
	\caption{Fits of the center (\ref{fig:furtherchirales1centerfit}) and right (\ref{fig:furtherchirales1rightfit}) branches of the lower levels of the $Q=\pm 1$ entanglement spectrum, corresponding to the $\ket{\bm{3}}$ or $\ket{\overline{\bm{3}}}$ primary sectors of the chiral $\mathrm{SU}(3)_1$ WZW theory. In each fit, 5 parameters, corresponding to the conserved integrals of the operators of the ``$\Phi_i(x)$ included'' column of Table \ref{table:operators}, are used to fit the 14 differences among the energies of the 15 $\mathrm{SU}(3)$ multiplets in the first 4 levels of the $\ket{\bm{3}}$ and $\ket{\overline{\bm{3}}}$ sectors.}
	\label{fig:furtherchiralesfit}
\end{figure}

\begin{figure}[H]
	\centering
	\subfloat[]{\label{fig:simultaneouschirales0leftfit} \includegraphics[scale=0.4]{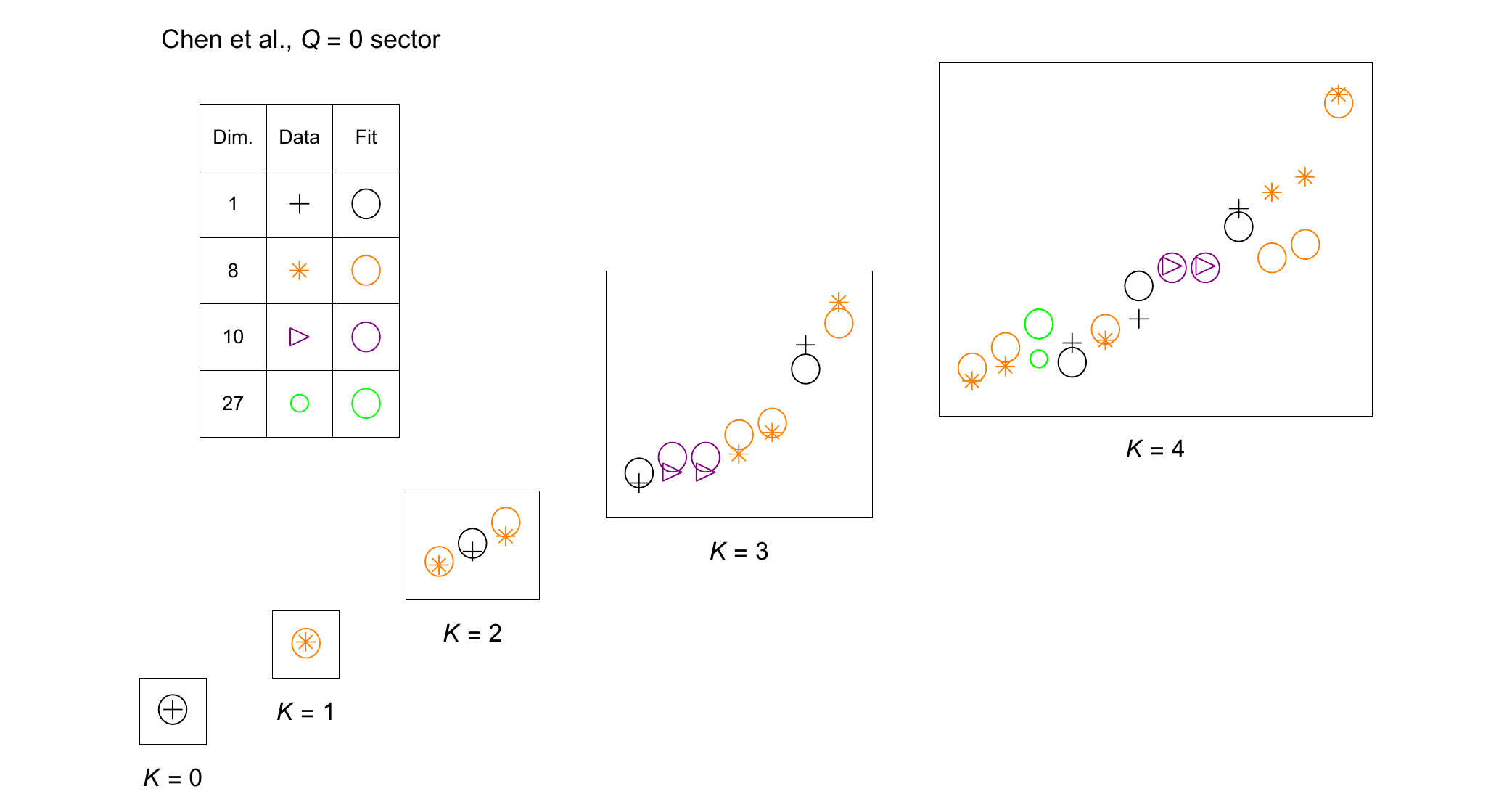}
	} \\
	\subfloat[]{\label{fig:simultaneouschirales1leftfit} \includegraphics[scale=0.4]{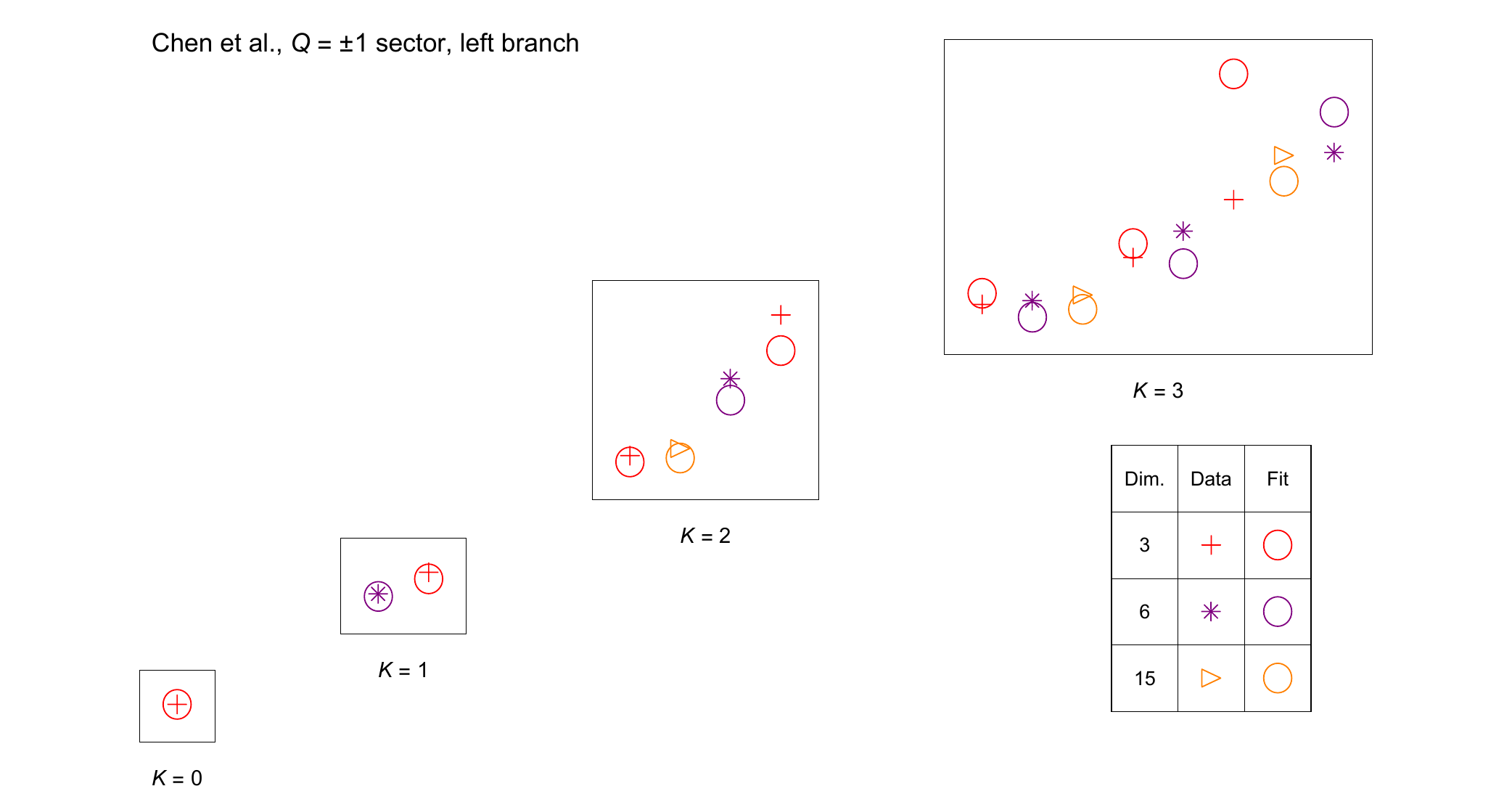}
	}
	\caption{Simultaneous fits of the lower levels of the $Q = 0$  (\ref{fig:simultaneouschirales0leftfit}) and $Q = \pm 1$  (\ref{fig:simultaneouschirales1leftfit}) sectors of the entanglement spectrum, incorporating the left branch of the $Q=\pm 1$ entanglement spectrum, respectively. The simultaneous pair of fits uses the same expression for the entanglement Hamiltonian evaluated in both the $\ket{\bm{1}}$ and $\ket{\bm{3}}$ or $\ket{\overline{\bm{3}}}$ primary sectors of the chiral $\mathrm{SU}(3)_1$ WZW theory. 
	In each fit, we use 6 parameters, corresponding to the 5 conserved integrals of the operators of the ``$\Phi_i(x)$ included'' column of Table \ref{table:operators}, along with a scale factor for the relative values of the $\tilde{\beta}_i$ in the two different sectors. These are used to simultaneously fit the 23 differences among the energies of the 24 $\mathrm{SU}(3)$ multiplets in the first 5 levels of the $\ket{\bm{1}}$ sector, and the 14 differences among the energies of the 15 $\mathrm{SU}(3)$ multiplets in the first 4 levels of the $\ket{\bm{3}}$ and $\ket{\overline{\bm{3}}}$ sectors.
	}
	\label{fig:simultaneouschiralesleftfit}
\end{figure}

\begin{figure}[H]
	\centering
	\subfloat[]{\label{fig:simultaneouschirales0centerfit} \includegraphics[scale=0.4]{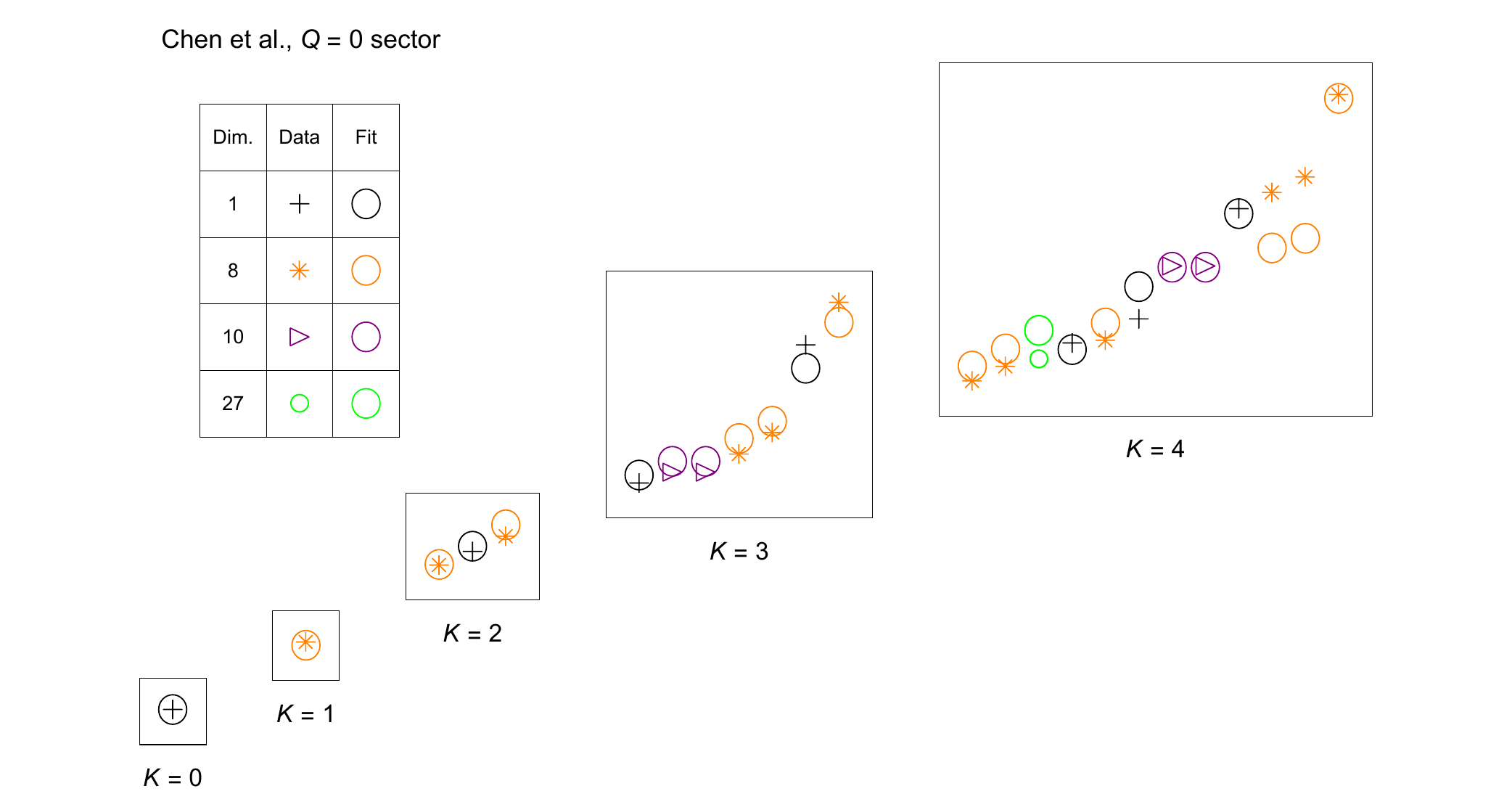}
	} \\
	\subfloat[]{\label{fig:simultaneouschirales1centerfit} \includegraphics[scale=0.4]{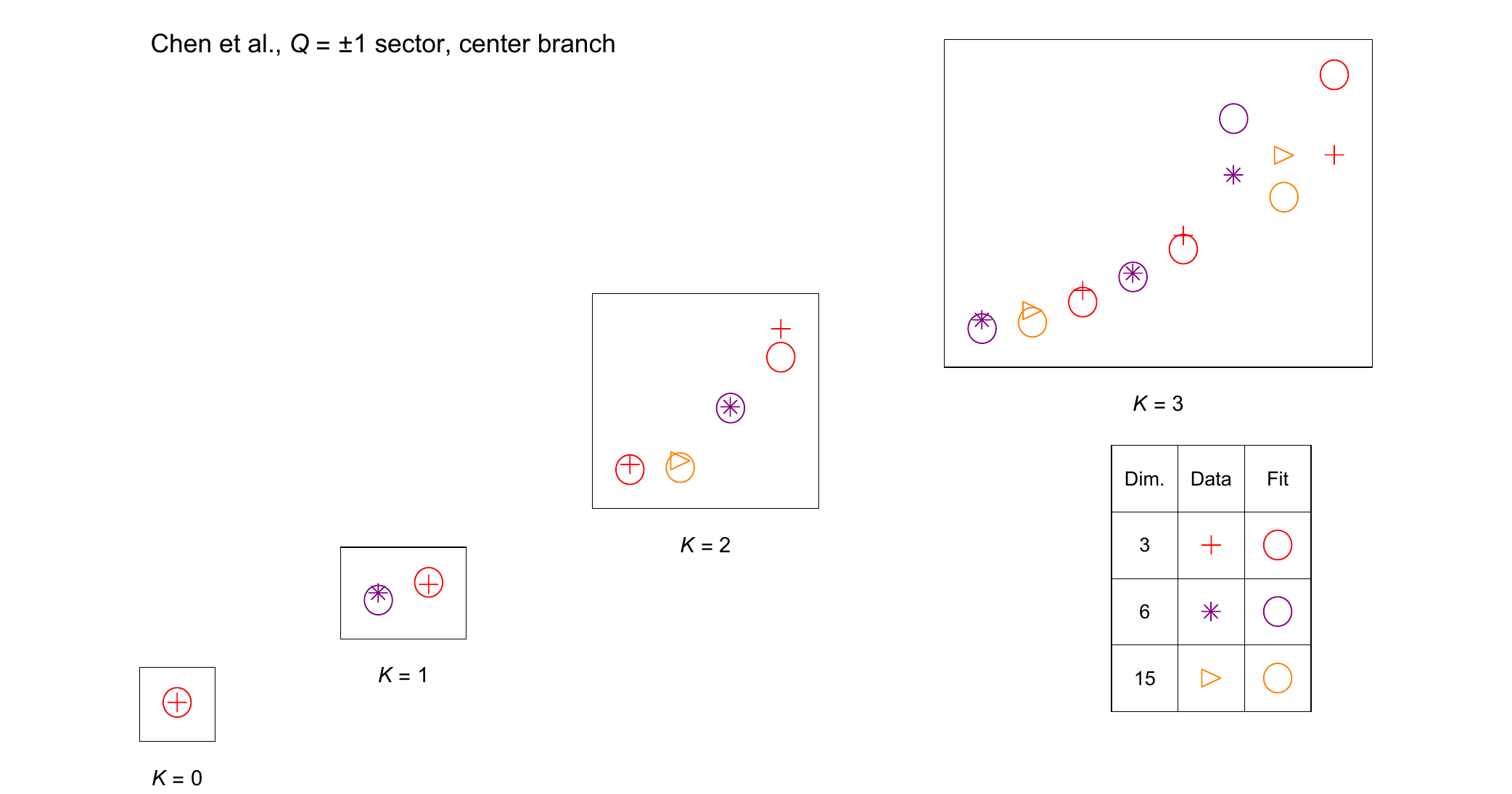}
	}
	\caption{Simultaneous fits of the lower levels of the $Q = 0$ (\ref{fig:simultaneouschirales0centerfit}) and $Q = \pm 1$  (\ref{fig:simultaneouschirales1centerfit}) sectors \\ of the entanglement spectrum, incorporating the center branch of the $Q=\pm 1$ entanglement spectrum, respectively. The simultaneous pair of fits uses the same expression for the entanglement Hamiltonian evaluated in both the $\ket{\bm{1}}$ and $\ket{\bm{3}}$ or $\ket{\overline{\bm{3}}}$ primary sectors of the chiral $\mathrm{SU}(3)_1$ WZW theory.
	In each fit, we use 6 parameters, corresponding to the 5 conserved integrals of the operators of the ``$\Phi_i(x)$ included'' column of Table \ref{table:operators}, along with a scale factor for the relative values of the $\tilde{\beta}_i$ in the two different sectors. These are used to simultaneously fit the 23 differences among the energies of the 24 $\mathrm{SU}(3)$ multiplets in the first 5 levels of the $\ket{\bm{1}}$ sector, and the 14 differences among the energies of the 15 $\mathrm{SU}(3)$ multiplets in the first 4 levels of the $\ket{\bm{3}}$ and $\ket{\overline{\bm{3}}}$ sectors.
	}
	\label{fig:simultaneouschiralescenterfit}
\end{figure}

\begin{figure}[H]
	\centering
	\subfloat[]{\label{fig:simultaneouschirales0rightfit} \includegraphics[scale=0.4]{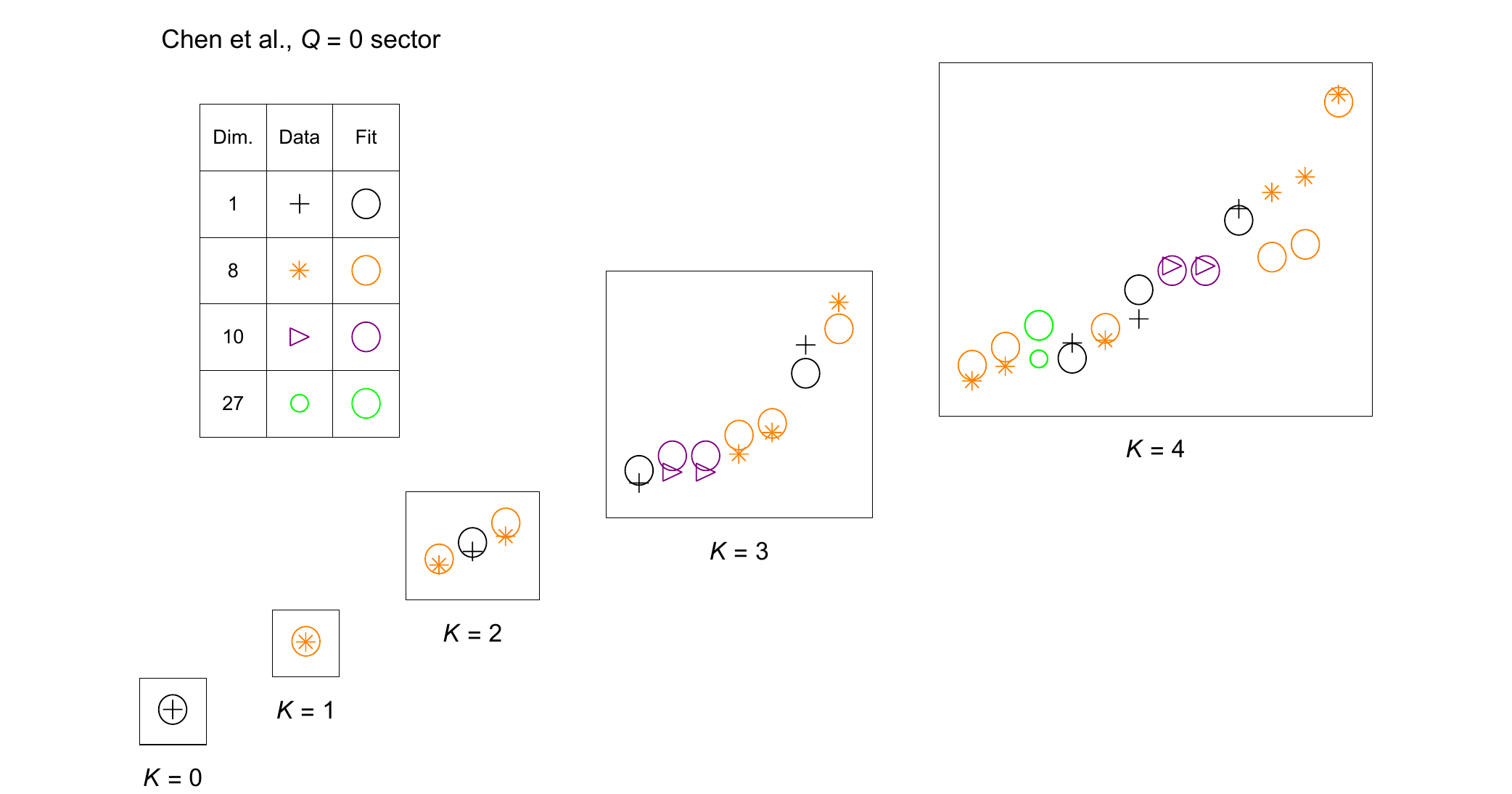}
	} \\
	\subfloat[]{\label{fig:simultaneouschirales1rightfit} \includegraphics[scale=0.4]{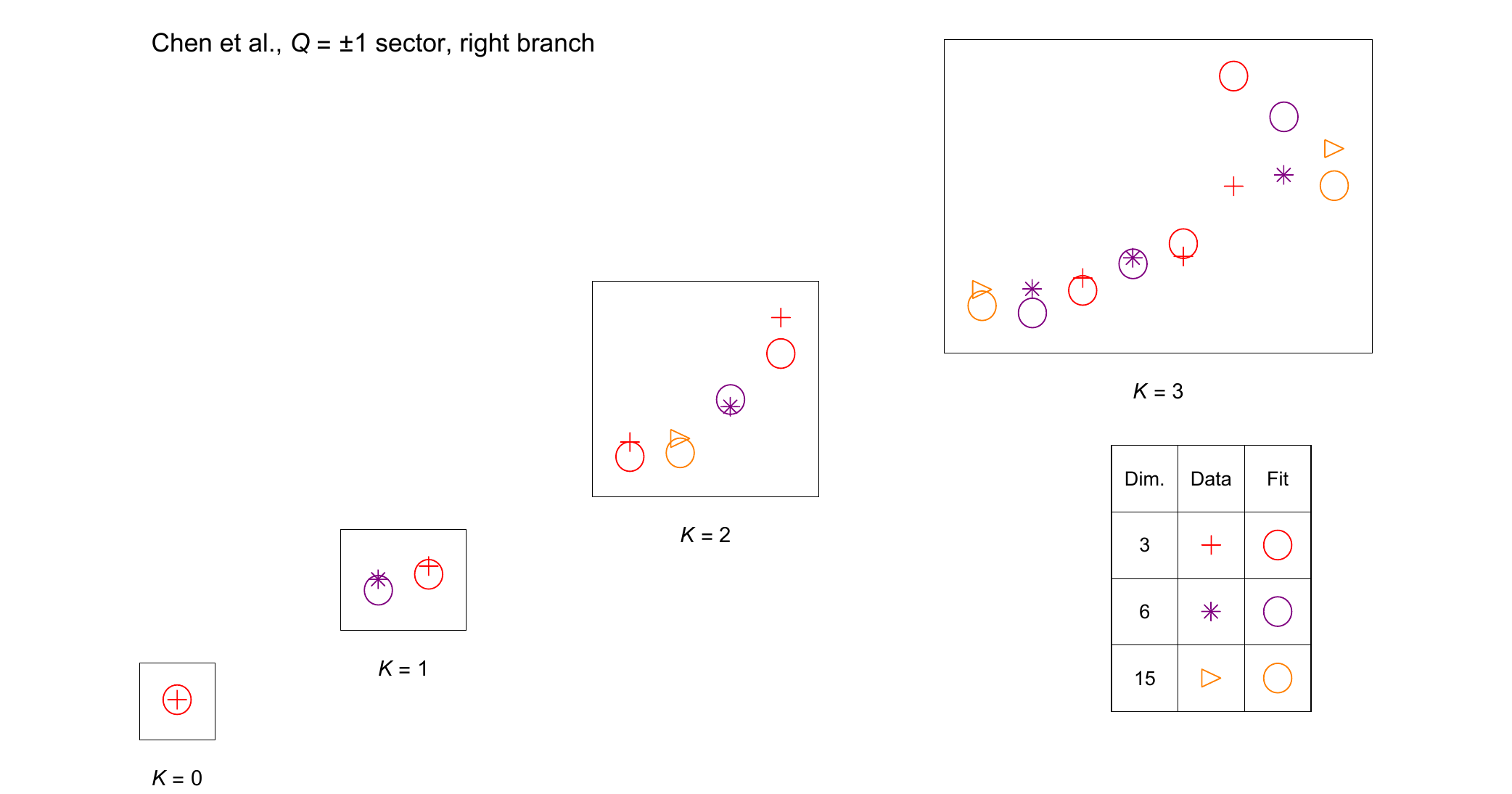}
	}
	\caption{Simultaneous fits of the lower levels of the $Q = 0$ (\ref{fig:simultaneouschirales0rightfit}) and $Q = \pm 1$  (\ref{fig:simultaneouschirales1rightfit}) sectors of the entanglement spectrum, incorporating the right branch of the $Q=\pm 1$ entanglement spectrum, respectively. The simultaneous pair of fits uses the same expression for the entanglement Hamiltonian evaluated in both the $\ket{\bm{1}}$ and $\ket{\bm{3}}$ or $\ket{\overline{\bm{3}}}$ primary sectors of the chiral $\mathrm{SU}(3)_1$ WZW theory. 
	In each fit, we use 6 parameters, corresponding to the 5 conserved integrals of the operators of the ``$\Phi_i(x)$ included'' column of Table \ref{table:operators}, along with a scale factor for the relative values of the $\tilde{\beta}_i$ in the two different sectors. These are used to simultaneously fit the 23 differences among the energies of the 24 $\mathrm{SU}(3)$ multiplets in the first 5 levels of the $\ket{\bm{1}}$ sector, and the 14 differences among the energies of the 15 $\mathrm{SU}(3)$ multiplets in the first 4 levels of the $\ket{\bm{3}}$ and $\ket{\overline{\bm{3}}}$ sectors.
	}
	\label{fig:simultaneouschiralesrightfit}
\end{figure}

In Sec.~\ref{sec:results}, we exhibited fits of our model for the entanglement spectrum to the entanglement spectrum of the PEPS of Sec.~\ref{sec:peps} in the $Q = 0$ sector, corresponding to the $\ket{\bm{1}}$ primary sector, as well as the right branch of the $Q = \pm 1$ sector, corresponding to the $\ket{\bm{3}}$ or $\ket{\overline{\bm{3}}}$ primary sectors. These fits were conducted separately. In this section, we exhibit further fits of our model for the $\ket{\bm{3}}$ or $\ket{\overline{\bm{3}}}$ primary sector to the center and right branches of the $Q = \pm 1$ sector of the data. These are found in Figure \ref{fig:furtherchiralesfit}. We also exhibit simultaneous fits of the $Q = 0$ sector and each of the three branches of the $Q = \pm 1$ sector, where we determine the $\beta_i$ in Eq.~\eqref{eq:lincomb} that lead to an entanglement Hamiltonian that simultaneously best fits the data in both the $\ket{\bm{1}}$ and the $\ket{\bm{3}}$ or $\ket{\overline{\bm{3}}}$ primary sectors. (Specifically, we allow the $\beta_i$ in the two sectors to differ by a constant scale factor, also fit.) These fits are found in Figures \ref{fig:simultaneouschiralesleftfit}-\ref{fig:simultaneouschiralesrightfit}. 


\section{Parameter Values of the Entanglement Spectrum Fits}
\label{app:params}

To perform the fits to the entanglement spectrum in a particular sector (either $\ket{\bm{1}}$ or $\ket{\bm{3}}$/$\ket{\overline{\bm{3}}}$), we select GGE parameters $\beta_i$ for our model [Eq.~\eqref{eq:lincomb}] that minimize the least squares fitting function
\begin{equation}
\label{eq:fitfunc}
R\left(\{\beta_i\}\right)=\sum_\ell \left[\xi^{\text{ES}}_\ell-\xi^{\text{model}}_\ell\left(\{\beta_i\}\right) \right]^2 W_\ell,
\end{equation}
where $\xi_\ell$ is the $\ell$th eigenvalue in a standard ordering of the eigenvalues of the entanglement spectrum in that sector (of the numerical entanglement spectrum data, in the case of $\xi^{\text{ES}}_\ell$, or of our model, in the case of $\xi^{\text{model}}_\ell$. $W_\ell$ is a normalized weight factor set to be inversely proportional to the number of multiplets in the descendant level containing $\xi_\ell$, so as to weight each descendant level equally in the fitting function. For the plots of Sec.~\ref{sec:results}, Figs.~\ref{fig:chirales0fit} and \ref{fig:chirales1fit}, and of Fig.~\ref{fig:furtherchiralesfit} in Appendix \ref{sec:furtherfits}, we fit the sectors separately, minimizing Eq.~\eqref{eq:fitfunc} for each of the $\ket{\bm{1}}$ and $\ket{\bm{3}}$/$\ket{\overline{\bm{3}}}$ sectors individually. On the other hand, for the simultaneous fits shown in Figs.~\ref{fig:simultaneouschiralesleftfit}-\ref{fig:simultaneouschiralesrightfit} in Appendix \ref{sec:furtherfits}, we instead minimize the sum of the fitting functions [Eq.~\eqref{eq:fitfunc}] for each of the $\ket{\bm{1}}$ and $\ket{\bm{3}}$/$\ket{\overline{\bm{3}}}$ sectors. 

One additional consideration is that the expressions $\tilde{H}^{(i)}$ we have for the integrals of motion (see Table \ref{table:modereps}) have their size-dependence divided out. Thus the actual parameters $\tilde{\beta}_i$ we calculate are size-dependent. We have
\begin{align}
\tilde{H}^{(i)} &= \left(\frac{\ell}{2\pi}\right)^{\Delta_i - 1}H^{(i)} \\
\tilde{\beta}_i &= \left(\frac{2\pi}{\ell}\right)^{\Delta_i - 1}\beta_i.
\end{align}
Thus, Eq.~\eqref{eq:lincomb} becomes
\begin{equation}
    \label{eq:lincombindep}
    H_{\text{entanglement}} = \tilde{\beta} \tilde{H}_{L} + \sum_{i=2}^\infty \tilde{\beta}_i \tilde{H}^{(i)}.
\end{equation}

In Table \ref{table:su31fitparams}, we exhibit the numerical values of the fitting parameters $\tilde{\beta}_i$ from Eq.~\eqref{eq:lincombindep} for the various fits performed in Sec.~\ref{sec:results} and Appendix \ref{sec:furtherfits}, along with the corresponding best fit value of the fitting function $R\left(\{\beta_i\}\right)$ of Eq.~\eqref{eq:fitfunc}. These have been normalized to remove an arbitrary factor of scale so that $\tilde{\beta}_1 = 1$.

\begin{table}[hbt]
	\centering
	\begin{tabular}{c|c|c|c|c|c|c||c}
		Fit & Figure & Sector & $\tilde{\beta}_3$ & $\tilde{\beta}_5$ & $\tilde{\beta}_6$ & $\tilde{\beta}_8$ & $R\left(\{\beta_i\}\right)$ \\
		\hline \hline
		 $Q = 0$ separate & Fig.~\ref{fig:chirales0fit} & $\ket{\bm{1}}$ & 0.239 & -0.00295 & -0.0167 & -0.0155 & 0.0633 \\
		 \hline
         Left $Q = \pm 1$ separate & Fig.~\ref{fig:chirales1fit} & $\ket{\bm{3}}$/$\ket{\overline{\bm{3}}}$ & 0.333 & -0.0136 & -0.0263 & -0.0327 & 0.00477 \\
		 \hline
         Center $Q = \pm 1$ separate & Fig.~\ref{fig:furtherchirales1centerfit} & $\ket{\bm{3}}$/$\ket{\overline{\bm{3}}}$ & 0.296 & -0.012 & -0.0172 & -0.0267 & 0.0163 \\
		 \hline
         Right $Q = \pm 1$ separate & Fig.~\ref{fig:furtherchirales1rightfit} & $\ket{\bm{3}}$/$\ket{\overline{\bm{3}}}$ & 0.214 & -0.00763 & -0.0172 & -0.0217 & 0.0158 \\
         \hline \hline
         \multirow{2}{*}{$Q = 0$ and left $Q = \pm 1$ simultaneous} & Fig.~\ref{fig:simultaneouschirales0leftfit} & $\ket{\bm{1}}$ & \multirow{2}{*}{0.179} & \multirow{2}{*}{-0.00104} & \multirow{2}{*}{-0.0151} & \multirow{2}{*}{-0.0205} & \multirow{2}{*}{0.187} \\
          & Fig.~\ref{fig:simultaneouschirales1leftfit} & $\ket{\bm{3}}$/$\ket{\overline{\bm{3}}}$ & & & & & \\
         \hline
         \multirow{2}{*}{$Q = 0$ and center $Q = \pm 1$ simultaneous} & Fig.~\ref{fig:simultaneouschirales0centerfit} & $\ket{\bm{1}}$ & \multirow{2}{*}{0.191} & \multirow{2}{*}{-0.00159} & \multirow{2}{*}{-0.0149} & \multirow{2}{*}{-0.0186} & \multirow{2}{*}{0.138} \\
          & Fig.~\ref{fig:simultaneouschirales1centerfit} & $\ket{\bm{3}}$/$\ket{\overline{\bm{3}}}$ & & & & &  \\
         \hline
         \multirow{2}{*}{$Q = 0$ and right $Q = \pm 1$ simultaneous} & Fig.~\ref{fig:simultaneouschirales0rightfit} & $\ket{\bm{1}}$ & \multirow{2}{*}{0.154} & \multirow{2}{*}{-0.00014} & \multirow{2}{*}{-0.0135} & \multirow{2}{*}{-0.0181} & \multirow{2}{*}{0.168} \\
          & Fig.~\ref{fig:simultaneouschirales1rightfit} & $\ket{\bm{3}}$/$\ket{\overline{\bm{3}}}$ &  &  &  &  &  \\
		\hline 
	\end{tabular}
	\caption{Normalized numerical values of the best-fit choices of $\tilde{\beta}_i$ (where $\tilde{\beta}_i$ is the parameter corresponding to the $i$th conserved quantity of Table \ref{table:modereps}) that were used to generate fits to the $\mathrm{SU}(3)_1$ data of Sec.~\ref{sec:results} and Appendix \ref{sec:furtherfits}, along with the associated value of the fitting function $R\left(\{\beta_i\}\right)$. Left, center and right $Q = \pm 1$ fits refer to the left, center, and right chiral branches found in the lower levels of the $Q = \pm 1$ entanglement spectrum (as defined in Sec.~\ref{sec:results}).}
	\label{table:su31fitparams}
\end{table}

\section{Additional information on the splittings observed in the non-chiral PEPS}
\label{app:nonchiralsplittings}

\begin{figure}[H]
	\centering
	\includegraphics[scale=0.5]{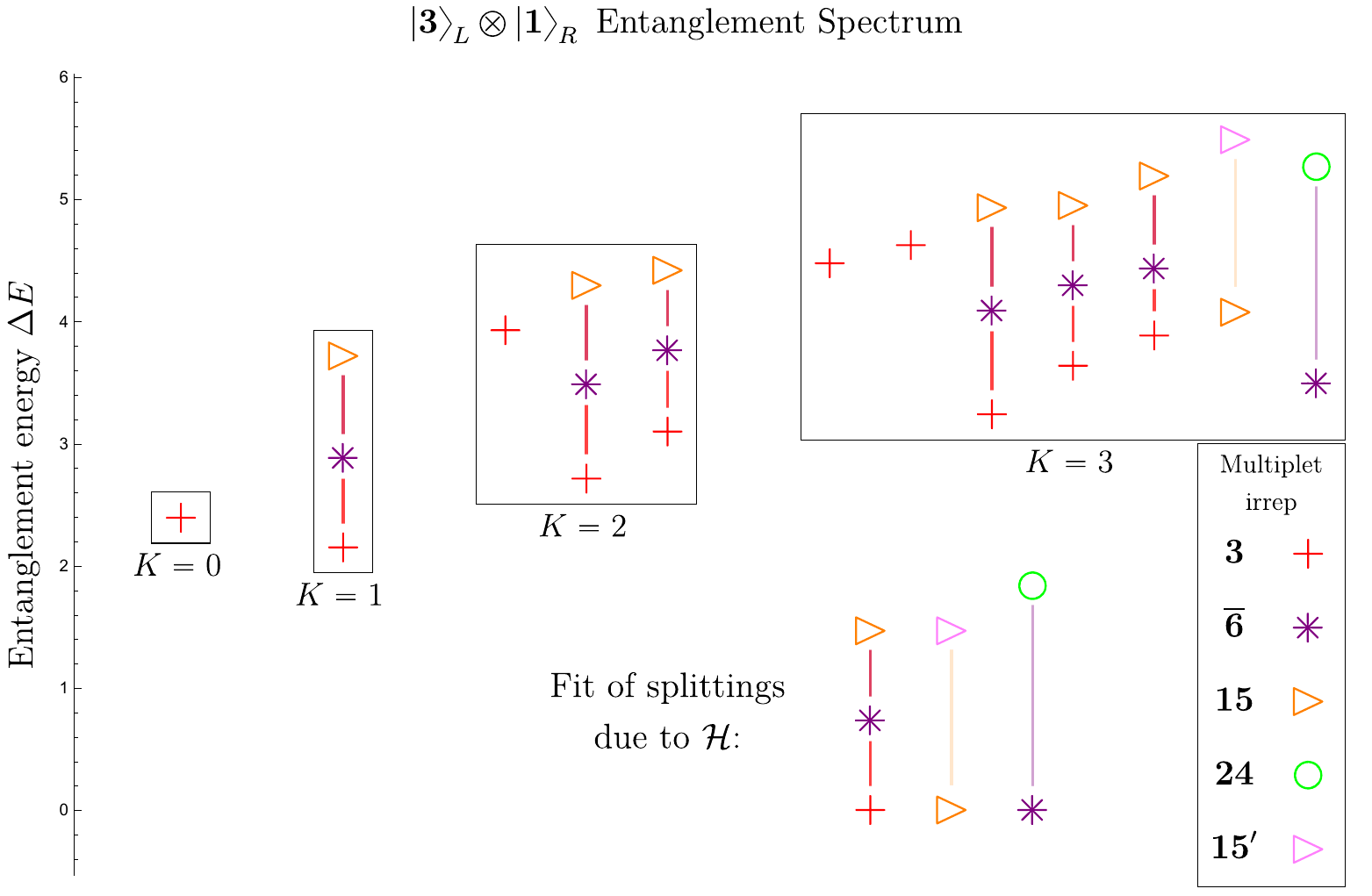}
	\caption{The entanglement spectrum of the non-chiral PEPS is shown in the $\ket{\bm{3}}_L\otimes\ket{\bm{1}}_R$ sector. [Recall that the $L$ subscript denotes a primary state in the ``fast'', left-moving ($L$) chiral branch. An $R$ subscript denotes a primary state in the ``slow'', right-moving ($R$) chiral branch.] The splittings between multiplets are depicted by the lines between them. The fit of the different sorts of these splittings with one overall parameter multiplied by the perturbation $\mathcal{H}^{(1)}$ of Eq.~\eqref{eq:jjbarperturbation}, that is, the fit of all the differences between states within the same tensor product making use of Eq.~\eqref{eq:deltas} with a single uniform parameter $\lambda_1$ used to fit all the data, is shown at the bottom for comparison. The multiplet content of this spectrum and the depicted splittings correspond to those enumerated in Table \ref{table:multipletbreakdownmaintext}. The horizontal separation of the data points within each box at level $K$ has been artificially added in order to more clearly show overlapping data.}
	\label{fig:nonchiralessplittings11}
\end{figure}

\begin{figure}[H]
	\centering
	\includegraphics[scale=0.5]{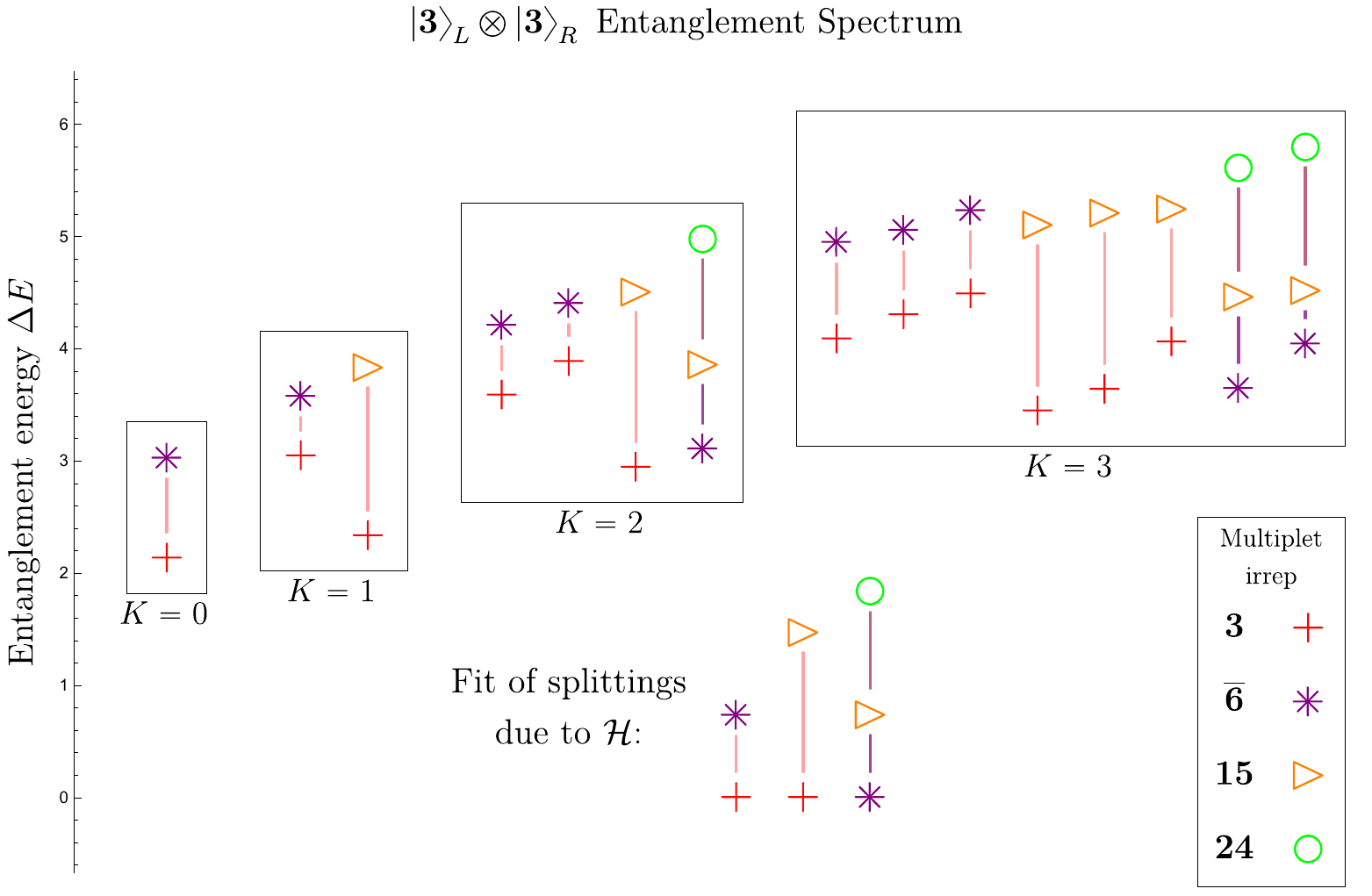}
	\caption{The entanglement spectrum of the non-chiral PEPS is shown in the $\ket{\bm{3}}_L\otimes\ket{\bm{3}}_R$ sector. [Recall that the $L$ subscript denotes a primary state in the ``fast'', left-moving ($L$) chiral branch. An $R$ subscript denotes a primary state in the ``slow'', right-moving ($R$) chiral branch.] The splittings between multiplets are depicted by the lines between them. The fit of the different sorts of these splittings with one overall parameter multiplied by the perturbation $\mathcal{H}^{(1)}$ of Eq.~\eqref{eq:jjbarperturbation}, that is, the fit of all the differences between states within the same tensor product making use of Eq.~\eqref{eq:deltas} with a single uniform parameter $\lambda_1$ used to fit all the data, is shown at the bottom for comparison. The horizontal separation of the data points within each box at level $K$ has been artificially added in order to more clearly show overlapping data.}
	\label{fig:nonchiralessplittings20}
\end{figure}

The complete table of the composition of the lowest four levels of entanglement spectrum of the non-chiral PEPS in three representative sectors, extending that for the $\ket{\bm{3}}_L\otimes\ket{\bm{\bar{3}}}_R$ given in Table \ref{table:multipletbreakdownmaintext}, is shown in Table \ref{table:multipletbreakdown}. Additional plots of entanglement spectrum splittings between states originating from the same tensor product of left-and right-chiral states in the non-chiral PEPS exhibited in Sec.~\ref{sec:nonchiralsplit} are also given. Such splittings in the $\ket{\bm{3}}_L\otimes\ket{\bm{\bar{3}}}_R$ sector are found in Fig.~\ref{fig:nonchiralessplittings02} in the main text, while two other sectors can be found in this Appendix: Fig.~\ref{fig:nonchiralessplittings11}, which displays the splittings found in the $\ket{\bm{3}}_L\otimes\ket{\bm{1}}_R$ sector entanglement spectrum, and Fig.~\ref{fig:nonchiralessplittings20}, which displays those of the $\ket{\bm{3}}_L\otimes\ket{\bm{3}}_R$ sector.

\begin{table}[H]
    \centering
    \begin{tabular}{c|c|c||c|x{1.7cm}|x{1.7cm}|x{1.7cm}|x{1.7cm}|x{1.7cm}}
        $(q,\phi)$ & \text{Fast ($L$)} & \text{Slow ($R$)} & \multicolumn{6}{c}{Multiplet content} \\
        \hline
        \hline
        \multirow{7}{*}{$(0,2)$} & \multirow{7}{*}{$\bm{3}$} & \multirow{7}{*}{$\overline{\bm{3}}$} & \multirow{2}{*}{Multiplets} & $\bm{3}\otimes\overline{\bm{3}}$& \multicolumn{2}{c|}{$\bm{3}\otimes\bm{6}$} & \multicolumn{2}{c}{$\bm{3}\otimes\overline{\bm{15}}$} \\
        &  &  &  & $=\bm{1}+\bm{8}$ & \multicolumn{2}{c|}{$=\bm{8}+\bm{10}$} & \multicolumn{2}{c}{$=\bm{8}+\overline{\bm{10}}+\bm{27}$}  \\
        \cline{4-9} 
        &  &  & $\Delta \mathcal{C}^{(2)}$ & 3 & \multicolumn{2}{c|}{3} & 3 & 2 \\
        \cline{4-9}
        &  &  & $K = 0$ & 1 & \multicolumn{2}{c|}{0} & \multicolumn{2}{c}{0} \\
        &  &  & $K = 1$ & 1 & \multicolumn{2}{c|}{1} & \multicolumn{2}{c}{0} \\
        &  &  & $K = 2$ & 2 & \multicolumn{2}{c|}{1} & \multicolumn{2}{c}{1} \\ 
        &  &  & $K = 3$ & 3 & \multicolumn{2}{c|}{3} & \multicolumn{2}{c}{2} \\
        \hline
        \hline
        \multirow{7}{*}{$(1,1)$} & \multirow{7}{*}{$\bm{3}$} & \multirow{7}{*}{$\bm{1}$} &
        \multirow{2}{*}{Multiplets} &
        $\bm{3}\otimes\bm{1}$ & \multicolumn{2}{c|}{$\bm{3}\otimes\bm{8}$} & $\bm{3}\otimes\bm{10}$ &  $\bm{3}\otimes\overline{\bm{10}}$ \\
        &  &  &  & $=\bm{3}$ & \multicolumn{2}{c|}{$=\bm{3}+\overline{\bm{6}}+\bm{15}$} & $=\bm{15}+\bm{15}'$ & $=\overline{\bm{6}}+\bm{24}$ \\
        \cline{4-9}
        &  &  & $\Delta \mathcal{C}^{(2)}$ & --- & 2 & 2 & 4 & 5 \\
        \cline{4-9}
        &  &  & $K=0$ & 1 & \multicolumn{2}{c|}{0} & 0 & 0 \\
        &  &  & $K=1$ & 0 & \multicolumn{2}{c|}{1} & 0 & 0 \\
        &  &  & $K=2$ & 1 & \multicolumn{2}{c|}{2} & 0 & 0 \\
        &  &  & $K=3$ & 1 & \multicolumn{2}{c|}{3} & 1 & 1 \\
        \hline
        \hline
        \multirow{7}{*}{$(2,0)$} & \multirow{7}{*}{$\bm{3}$} & \multirow{7}{*}{$\bm{3}$} &
        \multirow{2}{*}{Multiplets} &
        $\bm{3}\otimes\bm{3}$ & \multicolumn{2}{c|}{$\bm{3}\otimes\overline{\bm{6}}$} & \multicolumn{2}{c}{$\bm{3}\otimes\bm{15}$} \\
        &  &  &  & $=\overline{\bm{3}}+\bm{6}$ & \multicolumn{2}{c|}{$=\overline{\bm{3}}+\overline{\bm{15}}$} & \multicolumn{2}{c}{$=\bm{6}+\overline{\bm{15}}+\overline{\bm{24}}$} \\
        \cline{4-9}
        &  &  & $\Delta \mathcal{C}^{(2)}$ & 2 & \multicolumn{2}{c|}{4} & 2 & 3 \\
        \cline{4-9}
        &  &  & $K=0$ & 1 & \multicolumn{2}{c|}{0} & \multicolumn{2}{c}{0} \\
        &  &  & $K=1$ & 1 & \multicolumn{2}{c|}{1} & \multicolumn{2}{c}{0} \\
        &  &  & $K=2$ & 2 & \multicolumn{2}{c|}{1} & \multicolumn{2}{c}{1} \\
        &  &  & $K=3$ & 3 & \multicolumn{2}{c|}{3} & \multicolumn{2}{c}{2} \\
        \hline
        \hline
	\end{tabular}
\caption{The composition of the lowest four levels of the entanglement spectrum in three representative sectors is shown here, extending Table \ref{table:multipletbreakdownmaintext}, which only showed the data for the $\ket{\bm 3}_L \otimes \ket{\overline{\bm 3}}_R$ primary sector. The multiplet content comes in each case from the tensor product of the fast primary $\mathrm{SU}(3)$ irrep with the content of the slow primary sector indicated. The value of $\Delta \mathcal{C}^{(2)}$ is given for each adjacent pair of multiplets in each tensor product: this is proportional to their splitting due to the effect of $\mathcal{H}^{(1)}$. See Eq.~\eqref{eq:deltas}.}	
\label{table:multipletbreakdown}
\end{table}

\end{document}